\documentclass[11pt]{article}
\usepackage{graphicx,ifthen,color,algorithm,palatino}
\usepackage{amsmath,amsthm,amssymb,amsfonts,verbatim}
\usepackage{mathrsfs,accents,setspace,algorithm}
\newboolean{UnBlinded}
\newboolean{DoubleSpaced}
\setboolean{UnBlinded}{true}
\setboolean{DoubleSpaced}{false}
\def\bDscr{\pmb{\mathscr{D}}}
\def\fhat{{\widehat f}}

\def\fTrue{f_{\mbox{\tiny true}}}
\def\real{{\mathbb R}}
\def\naturalNumbers{{\mathbb N}}

\def\COMMONabv{\mbox{C}}
\def\bx{\boldsymbol{x}}
\def\bI{\boldsymbol{I}}
\def\bA{\boldsymbol{A}}
\def\logit{\mbox{logit}}
\def\expit{\mbox{expit}}
\def\bv{\boldsymbol{v}}
\def\bw{\boldsymbol{w}}
\def\bOmega{\boldsymbol{\Omega}}
\def\diag{\mbox{diag}}
\def\bomega{\boldsymbol{\omega}}
\def\bU{\boldsymbol{U}}
\def\bd{\boldsymbol{d}}
\def\bz{\boldsymbol{z}}

\def\bsigma{\boldsymbol{\sigma}}

\def\punder{\underline{p}}
\def\tr{\mbox{tr}}
\def\bib{\vskip12pt\par\noindent\hangindent=1 true cm\hangafter=1}
\def\bkappa{\boldsymbol{\kappa}}
\def\bC{\boldsymbol{C}}
\def\bD{\boldsymbol{D}}
\def\bV{\boldsymbol{V}}
\def\bs{\boldsymbol{s}}
\def\bL{\boldsymbol{L}}
\def\sigsqeps{\sigeps^2}
\def\sumin{\sum_{i=1}^n}
\def\bdelta{\boldsymbol{\delta}}
\def\btheta{\boldsymbol{\theta}}
\def\buhat{{\widehat \bu}}
\def\betatilde{{\widetilde\beta}}
\def\butilde{{\widetilde \bu}}
\def\simind{\stackrel{{\tiny \mbox{ind.}}}{\sim}}
\def\bbeta{\boldsymbol{\beta}}
\def\bgamma{\boldsymbol{\gamma}}
\def\by{\boldsymbol{y}}
\def\bbetatilde{{\widetilde\bbeta}}
\def\sigeps{\sigma_{\varepsilon}}
\def\bX{\boldsymbol{X}}
\def\bc{\boldsymbol{c}}
\def\thickarrow{\longleftarrow}
\def\be{\boldsymbol{e}}
\def\ba{\boldsymbol{a}}
\def\bb{\boldsymbol{b}}
\def\bSigma{\boldsymbol{\Sigma}}
\def\smhalf{{\textstyle{\frac{1}{2}}}}
\def\bmu{\boldsymbol{\mu}}
\def\bu{\boldsymbol{u}}
\def\bZ{\boldsymbol{Z}}
\def\bzero{\boldsymbol{0}}
\def\bone{\boldsymbol{1}}
\def\smalldot{\mbox{\fontsize{2mm}{5em}\selectfont{$\bullet$}}}
\newcommand*{\bdotover}[1]{\accentset{\mbox{\smalldot}}{#1}}
\newcommand*{\circover}[1]{\accentset{\circ}{#1}}
\newcommand{\xNonLinearForm}[2]{\bdotover{x}^{#1}_{#2}}
\newcommand{\xLinearForm}[2]{\circover{x}^{#1}_{#2}}
\newcommand{\bxNonLinearForm}[2]{\bdotover{\bx}^{#1}_{#2}}
\newcommand{\bxLinearForm}[2]{\circover{\bx}^{#1}_{#2}}

\def\myand{\&\ }
\def\sigsqSUBuj{\sigma^2_{\mbox{\scriptsize{$u$}\scriptsize{$j$}}}}
\def\aSUBuj{a_{u\,j}}
\def\bkappaInterior{\bkappa_{\mbox{\tiny inter.}}}

\def\bxLinOrigi{\bxLinearForm{\mbox{\tiny orig}}{i}}
\def\bxNonOrigi{\bxNonLinearForm{\mbox{\tiny orig}}{i}}
\def\yOrig{y^{\mbox{\tiny orig}}}
\def\bxLini{\bxLinearForm{\null}{i}}
\def\bxNoni{\bxNonLinearForm{\null}{i}}
\def\xLinji{\xLinearForm{\null}{ji}}
\def\xNonji{\xNonLinearForm{\null}{ji}}
\def\xNonOne{\xNonLinearForm{\null}{1}}
\def\xNoni{\xNonLinearForm{\null}{i}}
\def\xNonn{\xNonLinearForm{\null}{n}}
\def\bxNon{\bxNonLinearForm{\null}{\null}}
\def\bsigsqSUBu{\bsigma_{\mbox{\tiny{$u$}}}^2}
\def\baSUBu{\ba_{\mbox{\tiny{$u$}}}}

\def\bUC{\bU_{\mbox{\tiny{$C$}}}}
\def\bVC{\bV_{\mbox{\tiny{$C$}}}}
\def\bdC{\bd_{\mbox{\tiny{$C$}}}}
\def\bUD{\bU_{\mbox{\tiny{$D$}}}}
\def\bVD{\bV_{\mbox{\tiny{$D$}}}}
\def\bdD{\bd_{\mbox{\tiny{$D$}}}}
\def\bsD{\bs_{\mbox{\tiny{$D$}}}}
\def\sigsqSUBbeta{\sigma^2_{\mbox{\tiny{$\beta$}}}}
\def\sigsqSUBu{\sigma^2_{\mbox{\tiny{$u$}}}}
\def\sigsqSUBeps{\sigma^2_{\mbox{\tiny{$\varepsilon$}}}}
\def\Nwarm{N_{\mbox{\tiny warm}}}
\def\Nkept{N_{\mbox{\tiny kept}}}
\def\bCOS{\bC_{\mbox{\tiny OS}}}
\def\bZOS{\bZ_{\mbox{\tiny OS}}}
\def\bCcDR{\bC_{\mbox{\tiny cDR}}}

\def\bLOStocDR{\bL_{\mbox{\tiny OS.to.cDR}}}
\def\bCDR{\bC_{\mbox{\tiny DR}}}
\def\byOrig{\by^{\mbox{\tiny orig}}}
\def\bxLinOrigj{\bxLinearForm{\mbox{\tiny orig}}{j}}
\def\bxNonOrigj{\bxNonLinearForm{\mbox{\tiny orig}}{j}}
\def\bxLinOne{\bxLinearForm{\null}{1}}
\def\bxLindLin{\bxLinearForm{\null}{\dLin}}
\def\bxNonOne{\bxNonLinearForm{\null}{1}}
\def\bxNonOne{\bxNonLinearForm{\null}{1}}
\def\bxNondNon{\bxNonLinearForm{\null}{\dNon}}
\def\bxLinj{\bxLinearForm{\null}{j}}
\def\bxNonj{\bxNonLinearForm{\null}{j}}
\def\meanyOrig{\mbox{mean}(\byOrig)}
\def\stdevyOrig{\mbox{st.dev}(\byOrig)}
\def\meanxOrigLinj{\mbox{mean}(\bxLinOrigj)}
\def\meanxOrigNonj{\mbox{mean}(\bxNonOrigj)}
\def\stdevxOrigLinj{\mbox{st.dev.}(\bxLinOrigj)}
\def\stdevxOrigNonj{\mbox{st.dev.}(\bxNonOrigj)}
\def\yTone{\textbf{yT1}}
\def\yToneadj{\textbf{yT1}_{\mbox{\scriptsize{adj}}}}
\def\XTy{\textbf{XTy}}
\def\XTyadj{\textbf{XTy}_{\mbox{\scriptsize{adj}}}}
\def\ZTy{\textbf{ZTy}}
\def\XTX{\textbf{XTX}}
\def\ZTX{\textbf{ZTX}}
\def\ZTXj{\textbf{ZTX}^{\mbox{\scriptsize{$\langle j\rangle$}}}}
\def\ZTXjd{\textbf{ZTX}^{\mbox{\scriptsize{$\langle j'\rangle$}}}}
\def\ZTXjT{\textbf{ZTX}^{\mbox{\scriptsize{$\langle j\rangle$}}T}}

\def\ZTyj{\textbf{ZTy}^{\mbox{\scriptsize{$\langle j\rangle$}}}}
\def\ZTyadj{\textbf{ZTy}_{\mbox{\scriptsize{adj}}}}
\def\ZTyjadj{\textbf{ZTy}^{\mbox{\scriptsize{$\langle j\rangle$}}}_{\mbox{\scriptsize{adj}}}}
\def\ZTZjj{\textbf{ZTZ}^{\mbox{\scriptsize{$\langle j,j\rangle$}}}}
\def\ZTZjjd{\textbf{ZTZ}^{\mbox{\scriptsize{$\langle j,j'\rangle$}}}}

\def\ZTZ{\textbf{ZTZ}}
\def\pDens{\mathfrak{p}}
\def\qDens{\mathfrak{q}}

\def\bgammaSUBbeta{\bgamma_{\mbox{\tiny{$\beta$}}}}
\def\gammaSUBbeta{\gamma_{\mbox{\tiny{$\beta$}}}}
\def\gammaSUBu{\gamma_{\mbox{\tiny{$u$}}}}
\def\bgammaSUBu{\bgamma_{\mbox{\tiny{$u$}}}}
\def\dNon{d_{\bullet}}
\def\dLin{d_{\circ}}
\def\gammaSUBuj{\gamma_{u\,j}}

\def\aSUBeps{a_{\varepsilon}}
\def\sSUBeps{s_{\varepsilon}}
\def\aSUBbeta{a_{\mbox{\tiny{$\beta$}}}}
\def\rhoSUBbeta{\rho_{\mbox{\tiny{$\beta$}}}}
\def\rhoSUBu{\rho_{\mbox{\tiny{$u$}}}}
\def\gammaSUBbetaj{\gamma_{\mbox{\tiny{$\beta$}\scriptsize{$j$}}}}
\def\gammaSUBbetaOne{\gamma_{\mbox{\tiny{$\beta$}\scriptsize{$1$}}}}
\def\gammaSUBbetaEnd{\gamma_{\mbox{\tiny{$\beta$,}\scriptsize{$\dLin+\dNon$}}}}
\def\sigmaSUBbeta{\sigma_{\mbox{\tiny{$\beta$}}}}
\def\sigmaSUBu{\sigma_{\mbox{\tiny{$u$}}}}
\def\bSUBbetaj{b_{\mbox{\tiny{$\beta$}\scriptsize{$j$}}}}
\def\bSUBbetaOne{b_{\mbox{\tiny{$\beta$}\scriptsize{$1$}}}}
\def\bSUBbetaLast{b_{\mbox{\tiny{$\beta$}\scriptsize{$(\dLin+\dNon)$}}}}
\def\gammaSUBuj{\gamma_{\mbox{\scriptsize{$u$}\scriptsize{$j$}}}}
\def\bSUBuj{b_{\mbox{\scriptsize{$u$}\scriptsize{$j$}}}}
\def\abeta{a_{\mbox{\tiny{$\beta$}}}}
\def\sSUBbeta{s_{\mbox{\tiny{$\beta$}}}}
\def\CSUBK{C_{\mbox{\tiny{$K$}}}}
\def\sSUBu{s_{\mbox{\tiny{$u$}}}}

\def\bwZj{\bw_{\mbox{\tiny $\bZ$}j}}
\def\bSUBbeta{b_{\mbox{\tiny{$\beta$}}}}
\def\bSUBu{b_{\mbox{\tiny{$u$}}}}
\def\bbSUBbeta{\bb_{\mbox{\tiny{$\beta$}}}}
\def\bbSUBu{\bb_{\mbox{\tiny{$u$}}}}
\def\bUOmega{\bU_{\mbox{\tiny{$\bOmega$}}}}
\def\bdOmega{\bd_{\mbox{\tiny{$\bOmega$}}}}

\def\bbetaCurr{\bbeta^{\mbox{\tiny{curr}}}}
\def\bujCurr{\bu_j^{\mbox{\tiny{curr}}}}
\def\bujdCurr{\bu_{j'}^{\mbox{\tiny{curr}}}}
\def\butildejCurr{\butilde_j^{\mbox{\tiny{curr}}}}
\def\butildejdCurr{\butilde_{j'}^{\mbox{\tiny{curr}}}}
\def\gammaSUBujd{\gamma_{u\,j'}}

\def\gammaSUBujCurr{\gamma_{\mbox{\tiny{$u$}}\mbox{\scriptsize{$j$}}}^{\mbox{\tiny{curr}}}}
\def\gammaSUBujdCurr{\gamma_{\mbox{\tiny{$u$}}\mbox{\scriptsize{$j'$}}}^{\mbox{\tiny{curr}}}}

\def\gothicc{\mathfrak{c}}
\def\omegaAAA{\omega_1}
\def\omegaBBB{\omega_2}
\def\omegaCCC{\bomega_3}
\def\omegaDDD{\omega_4}
\def\omegaEEE{\omega_5}
\def\omegaFFF{\bomega_6}
\def\omegaGGG{\bomega_7}
\def\omegaHHH{\bomega_8}
\def\omegaIII{\omega_9}
\def\omegaJJJ{\bomega_{10}}
\def\omegaKKK{\omega_{11}}
\def\omegaLLL{\omega_{12}}
\def\omegaMMM{\bomega_{13}}
\def\omegaNNN{\bomega_{14}}
\def\omegaOOO{\omega_{15}}
\def\omegaPPP{\bomega_{16}}
\def\omegaQQQ{\omega_{17}}
\def\omegaRRR{\bomega_{18}}
\def\omegaSSS{\omega_{19}}
\def\omegaTTT{\bomega_{20}}
\begin{document}

\ifthenelse{\boolean{DoubleSpaced}}{\setstretch{1.5}}{}

\vskip5mm
\centerline{\Large\bf Bayesian Generalized Additive Model Selection}
\vskip1mm
\centerline{\Large\bf Including a Fast Variational Option}
\vskip5mm
\centerline{\normalsize\sc By Virginia X. He and Matt P. Wand}
\vskip5mm
\centerline{\textit{University of Technology Sydney}}
\vskip6mm
\centerline{25th September, 2023}

\vskip6mm

\centerline{\large\bf Abstract}
\vskip2mm

We use Bayesian model selection paradigms, such as group least absolute shrinkage 
and selection operator priors, to facilitate generalized additive model selection. 
Our approach allows for the effects of continuous predictors to be categorized as 
either zero, linear or non-linear. Employment of carefully tailored auxiliary 
variables results in Gibbsian Markov chain Monte Carlo schemes for practical 
implementation of the approach. In addition, mean field variational algorithms with 
closed form updates are obtained. Whilst not as accurate, this fast variational option
enhances scalability to very large data sets. A package in the \textsf{R}
language aids use in practice.

\vskip3mm
\noindent
\textit{Keywords:} Markov chain Monte Carlo; mean field variational Bayes; 
nonparametric regression; \textsf{R} package; scalable methodology.

\section{Introduction}\label{sec:intro}

Generalized additive models offer attractive solutions to the problem of obtaining
parsimonious, flexible and interpretable regression fits when faced with, potentially,
large numbers of candidate predictors (e.g. Hastie \myand Tibshirani, 1990; Wood, 2017).
Generalized additive models methodology and software is into its fourth decade. Nevertheless, 
principled, scalable and reliable selection of a model still has room for improvement. 
The version of the problem treated here is that where each candidate
predictor is categorized into one of three classes: having zero effect,
having a linear effect or having a non-linear effect on the mean response.
We provide new and effective solutions to the problem by employing recent developments
in Bayesian model selection and Bayesian computing. An accompanying package in the
\textsf{R} language (\textsf{R} Core Team, 2023) allows immediate use of our new methodology.

Several approaches to the three-category generalized additive model selection problem have
been proposed, such those in Shively \textit{et al.} (1999), Ravikumar \textit{et al.} (2009), 
Reich \textit{et al.} (2009), Scheipl \textit{et al.} (2012) and Chouldechova \myand Hastie (2015).
Our approach is inspired and closely tied to that of Chouldechova \myand Hastie (2015) which has
the advantages of excellent scalability and an accompanying \textsf{R} package (Chouldechova \myand
Hastie, 2018). Key features of the Chouldechova \myand Hastie (2015) approach are: use of the 
group least absolute shrinkage and selection operator (LASSO), Demmler-Reinsch spline bases, 
regularization paths and cross-validatory selection of the regularization parameter. Both Gaussian 
and binary response cases are supported. Instead of the path and cross-validation aspects, we embed
their infrastructure into a Bayesian graphical model and invoke Bayesian principles for model selection.
Simulation results point to superior three-category model selection. Other advantages of our
Bayesian approaches are being able to traverse a bigger model sparse compared with the 
regularization path approach and avoiding the practical difficulties associated with
finding cross-validation minima.

Once a Bayesian version of the Chouldechova \myand Hastie (2015) model is specified, a
pertinent challenge is tractability of Markov chain Monte Carlo and mean field variational
Bayes approaches to approximate inference. We achieve this via the introduction of appropriate
auxiliary variables. The binary response case benefits from the Albert \myand Chib (1993)
auxiliary variable approach for probit links. The resultant graphical models are such
that all full conditional distributions have standard forms. As a consequence, Markov
chain Monte Carlo sampling is Gibbsian and the mean field variational Bayes have closed forms
-- both of which depend only on sufficient statistics of the input data. Combined with the orthogonality 
advantages of Demmler-Reinsch spline bases, the resultant fitting and inference is relatively
fast and scales well to large data sets.

A simulation study shows that the new Bayesian approaches offer improved performance in terms
of classification of effect types as being either zero, linear or non-linear, compared with 
that of Chouldechova \myand Hastie (2015). They also shown to perform well in comparison with
the Bayesian approach of Scheipl \textit{et al.} (2012), but are considerably faster.

The \textsf{R} package that accompanies this article's methodology is named 
\textsf{gamselBayes} (He \myand Wand, 2023). In Section \ref{sec:performance} 
we compare its performance with two other \textsf{R} packages: \textsf{gamsel} 
(Chouldechova \myand Hastie, 2022) and \textsf{spikeSlabGAM} (Scheipl, 2022)
which also provide three-category model
selection for generalized additive models. Note that there are many other 
\textsf{R} packages concerned with generalized additive model analysis, 
some of which employ versions of the LASSO-type approach used by \textsf{gamselBayes}.
Examples of such packages are \textsf{BayesX} (Umlauf, Kneib \myand Klein, 2023), 
\textsf{bamlss} (Umlauf \textit{et al.}, 2023) and \textsf{bmrs} (B\"urkner, 2022).

Descriptions of our models and their conversion to computation-friendly forms
are given in Section \ref{sec:modelDescrip}. Algorithms for practical fitting 
and model selection are listed in Section \ref{sec:pracFitSel}. We also point
to the \textsf{R} package, \textsf{gamselBayes}, that allows easy and immediate access
to the new methodology for users of the \textsf{R} language. Section \ref{sec:performance} 
assesses performance of the new approaches in comparison with existing approaches with 
having similar aims. Applications to actual data are illustrated in Section \ref{sec:dataIllus}.
We close with some concluding remarks in Section \ref{sec:conclud}.

\section{Model Description}\label{sec:modelDescrip}

The original input data are as follows:
$$(\bxLinOrigi,\bxNonOrigi,\yOrig_i),\quad 1\le i\le n,$$
where, for each $1\le i\le n$, $\bxLinOrigi$ denotes a $\dLin\times1$ 
vector of predictors that can only enter the model linearly
(e.g. binary predictors) and $\bxNonOrigi$ denotes a $\dNon\times1$ 
vector of continuous predictors that can enter the model
either linearly or non-linearly. For Bayesian fitting and inference
we work with standardized versions of the data.  This has advantages
such as the methodology being independent of units of measurement for fixed
hyperparameter settings and improved numerical stability. Algorithm \ref{alg:suffStats}
in Section \ref{sec:preProc} provides the operational details of the
standardization process. The full data to be used for fitting and model selection 
are
$$(\bxLini,\bxNoni,y_i),\quad 1\le i\le n,$$
where $\bxLini$ and $\bxNoni$ are standardized data versions of $\bxLinOrigi$
and $\bxNonOrigi$. Also, for the continuous response case the $y_i$ are the standardized
response data. In the binary response case the $y_i$ are not pre-processed and 
remain as values in $\{0,1\}$. For each $1\le i\le n$ let
$$\xLinji\equiv\mbox{the $j$th entry of $\bxLini$},\quad 1\le j\le\dLin
\quad\mbox{and}\quad
\xNonji\equiv\mbox{the $j$th entry of $\bxNoni$},\quad 1\le j\le\dNon.
$$

Generalized additive models involve linear predictors $\eta_i$, $1\le i\le n$, 
having the generic forms
\begin{equation}
\eta_i\equiv\beta_0+\sum_{j=1}^{\dLin}\beta_j\xLinji+\sum_{j=1}^{\dNon}f_j(\xNonji),
\label{eq:BarneyRubble}
\end{equation}
where $\beta_0,\ldots,\beta_{\dLin}$ are the coefficients of linear components
and the $f_j$ are smooth real-valued functions over an interval containing the $\xNonji$ data.

\subsection{Matrix Notation}

For any column vector $\ba$ we let $\Vert\ba\Vert\equiv(\ba^T\ba)^{1/2}$ denote the Euclidean norm of $\ba$ 
and $\ba_{-j}$ denote the column vector with the $j$th entry of $\ba$ omitted. 
If $\bb$ is a column vector having the same number of rows as $\ba$ then $\ba\odot\bb$  and $\ba\big/\bb$ are,
respectively, the column vectors formed from $\ba$ and $\bb$ by obtaining element-wise 
products and quotients. For any square matrix $\bA$ we let $\mbox{diagonal}(\bA)$ denote
the column vector containing the diagonal entries of $\bA$.

\subsection{Distributional Definitions}

Table \ref{tab:distribs} lists all distributions used in this
article. In particular, the parametrizations of the 
corresponding density functions and probability functions 
are provided.  In this table, and throughout this article, 
$\Gamma(x)\equiv\int_0^{\infty} t^{x-1}e^{-t}\,dt$ is the gamma function
and $\Phi$ denotes the $N(0,1)$ cumulative distribution function.

\begin{table}[ht]
\begin{center}
\begingroup
\setlength{\tabcolsep}{2pt}
\begin{tabular}{lll}
\hline
distribution     & density/probability function in $x$  & abbreviation \\[0.1ex]
\hline\\[-0.9ex]
Bernoulli            &$\wp^x(1-\wp)^{1-x};\quad x=0,1;$\ $0<\wp<1$   
&  $\mbox{Bernoulli}(\wp)$      \\[3.5ex]
Multivariate  & $|2\pi\bSigma|^{-1/2}
\,\exp\{-\smhalf(\bx-\bmu)^T$
& $N(\bmu,\bSigma)$ \\
 Normal & $\null\qquad\qquad\qquad\qquad\times\bSigma^{-1}(\bx-\bmu)\}$  &    \\[3.5ex]
Inverse Gamma            &$\displaystyle{\frac{\lambda^{\kappa}\,x^{-\kappa-1}e^{-\lambda/x}}
{\Gamma(\kappa)}};\quad  x>0;\  \kappa,\lambda>0$   
&  $\mbox{Inverse-Gamma}(\kappa,\lambda)$      \\[3.5ex]
Inverse Gaussian        &
$\displaystyle{\frac{\sqrt{\lambda}\exp\left\{\displaystyle{\frac{-\lambda(x-\mu)^2}{2\mu^2 x}}\right\}}
{\sqrt{2\pi x^3}}}
;\ x>0;\  \mu,\lambda>0$   
&  $\mbox{Inverse-Gaussian}(\mu,\lambda)$      \\[3.5ex]
Beta        &
$\displaystyle{\frac{\Gamma(\alpha+\beta)x^{\alpha-1}(1-x)^{\beta-1}}
{\Gamma(\alpha)\Gamma(\beta)}}
;\ 0<x<1;$   
&  $\mbox{Beta}(\alpha,\beta)$      \\[0ex]
&$\null\qquad\qquad\qquad\qquad\qquad\qquad\alpha,\beta>0$  &    \\[3.5ex]
Half-Cauchy&$\displaystyle{\frac{2}{\pi\sigma((x/\sigma)^2+1)}};\quad  
x>0;\  \sigma>0$ & $\mbox{Half-Cauchy}(\sigma)$ \\[3.5ex]
$\mbox{Truncated-Normal}_+$& $\displaystyle{\frac{\exp\{-(x-\mu)^2/(2\sigma^2)\}}
{\Phi(\mu/\sigma)\sqrt{2\pi\sigma^2}}}; x>0;\  \sigma>0$    
&  $\mbox{Truncated-Normal}_+(\mu,\sigma^2)$ \\[3.5ex]
\hline
\end{tabular}
\endgroup
\caption{\textit{Distributions used in this article and their
corresponding density/probability functions.}}
\label{tab:distribs}
\end{center}
\end{table}

\subsection{Model for a Smooth Function}\label{sec:fjFuncs}

Let $\xNonOne,\ldots,\xNonn$ be a typical continuous predictor data sample. The corresponding
smooth function model takes the form
\begin{equation}
f(\xNoni)\equiv\beta\,\xNoni+\sum_{k=1}^K u_k z_k(\xNoni),\quad 1\le i\le n,
\label{eq:PunchAndJudy}
\end{equation}
for coefficients $\beta$ and $\bu\equiv(u_1,\ldots,u_K)$. Here $\{z_k(\cdot):1\le k\le K\}$
is an appropriate spline basis over an interval containing the $\xNoni$ data. 
In accordance with the set-up of Chouldechova \myand Hastie (2015), we choose the spline basis
to have orthogonality properties and lead to computational speed-ups. These properties 
can be explained succinctly in matrix algebraic terms. Define
$$\bxNon\equiv\mbox{the $n\times1$ vector with $i$th entry $\xNoni$}\ \mbox{and}\ 
\bZ\equiv\mbox{the $n\times K$ matrix having $(i,k)$ entry $z_k(\xNoni)$.}$$
Then we construct $\bZ$ to satisfy 
\begin{equation}
\bZ^T\bone_n=\bZ^T\bxNon=\bzero_{K}\quad\mbox{and}\quad\bZ^T\bZ\ \mbox{is a diagonal matrix}.
\label{eq:CurryAndChips}
\end{equation}
Spline bases satisfying (\ref{eq:CurryAndChips}) are referred to as having a \emph{Demmler-Reinsch} form.
In addition, we scale the columns of $\bZ$ so that the right-hand side of (\ref{eq:PunchAndJudy}) 
has mixed model representations of the form
\begin{equation}
\bxNon\beta+\bZ\bu\quad\mbox{where $\bu$ is a random vector having density function}\quad
\pDens(\bu)=h(\Vert\bu\Vert)
\label{eq:FortitudeValley}
\end{equation}
for some scalar-valued function $h$. In other words, we apply linear transformations to ensure
that the distribution of $\bu$ has spherical, rather than ellipsoidal, contours.
For ordinary generalized additive model fitting, as opposed
to selection, the most common choice of $h$ is 
$h(x)=(2\pi\sigma_u^2)^{-K/2}\exp\big\{-x^2/(2\sigma_u^2)\big\}$, for some $\sigma_u>0$, 
which corresponds to the spline coefficients model taking the form 
\begin{equation}
\bu|\sigma^2_u\sim N(0,\sigma_u^2\bI).
\label{eq:uNormal}
\end{equation}
For the generalized additive model selection, (\ref{eq:uNormal}) should be replaced by
an appropriate sparse signal prior distribution. Section \ref{sec:splineCoeffs} provides
full details on this modelling aspect. 

There are various ways in which $\{z_k(\cdot):1\le k\le K\}$ can be set up
so that (\ref{eq:CurryAndChips}) and (\ref{eq:FortitudeValley}) are satisfied.
In this article we follow the constructions laid out in Section 4 of Wand \myand Ormerod (2008) 
and Algorithm 1 of Ngo \myand Wand (2004). The full details are given in Section \ref{sec:ZcDR}.
of the supplement. We use the descriptor \emph{canonical Demmler-Reinsch basis} for this type 
of spline basis.

\subsection{Model for a Linear Coefficient}\label{sec:modLinCoeff}

Let $\beta$ denote a generic linear coefficient. We impose the following family of distributions
on $\beta$:
\begin{equation}
\pDens(\beta|\sigmaSUBbeta,\rhoSUBbeta)=\rhoSUBbeta(2\sigmaSUBbeta)^{-1}\exp\big(-|\beta|/\sigmaSUBbeta\big)
+(1-\rhoSUBbeta)\delta_0(\beta)
\label{eq:TwinCities}
\end{equation}
for parameters $\sigmaSUBbeta>0$ and $0\le\rhoSUBbeta\le1$. Here $\delta_0$ denotes the Dirac delta
function at zero. We call (\ref{eq:TwinCities}) the \emph{Laplace-Zero} family of distributions,
since it is a ``spike-and-slab'' mixture of a Laplace density function and a point mass at zero.
(e.g. Mitchell \myand Beauchamp, 1988).

The $\rhoSUBbeta=1$ version of (\ref{eq:TwinCities}) corresponds to the Bayesian Lasso approach 
of Park \myand Casella (2008). However, as pointed out there, Bayes estimation does not
lead to sparse fits for the purely Laplace prior situation. The addition of a point mass
at zero has the attraction of posterior distributions also having this feature
and sparse Bayes-type fits. This aspect is exploited in Section \ref{sec:modSelecStrat} for
principled model selection strategies.

The scale parameter in (\ref{eq:TwinCities}) has the prior distribution:
$$\sigmaSUBbeta\sim\mbox{Half-Cauchy}(\sSUBbeta)$$
for a hyperparameter $\sSUBbeta>0$. Gelman (2006) provides
justification for the imposition of a Half Cauchy prior on scale parameters 
such as $\sigmaSUBbeta$. The mixture parameter $\rhoSUBbeta$ is treated
as a hyperparameter.

Many alternatives to (\ref{eq:TwinCities}) for Bayesian model selection
have been proposed and studied. The overarching goal is the achievement of
sparse solutions, as is the case for frequentist LASSO-type approaches,
according to Bayesian fitting paradigms. The most common approach is to use  
``spike-and-slab'' priors, for which (\ref{eq:TwinCities}) is a special case, and 
involves mixing a symmetric zero mean continuous random variable with either a point 
mass at zero or another continuous random variable that is highly concentrated around zero. 
Key references include Lempers (1971), Mitchell \myand Beauchamp (1988), 
George \myand McCulloch (1993) and Ishwaran \myand Rao (2005).
Alternative approaches involve a single continuous distributional form, rather than a mixture, that
is sharply peaked at the origin and heavy-tailed. Examples include Park \myand Casella (2008), 
Carvalho \textit{et al.} (2010) and Griffin \myand Brown (2011). Bhadra \textit{et al.} (2019) 
compare and contrast both types of approaches.

\subsection{Model for a Spline Coefficients Vector}\label{sec:splineCoeffs}

Let $\bu$ denote a $K\times1$ spline coefficient vector. We impose the following 
family of distributions on $\bu$:
\begin{equation}
\pDens(\bu|\sigmaSUBu,\rhoSUBu)=\rhoSUBu(\CSUBK\sigmaSUBu)^{-1}
\exp\big(-\Vert\bu\Vert/\sigmaSUBu\big)
+(1-\rhoSUBu)\bdelta_0(\bu)
\label{eq:DontArgue}
\end{equation}
for parameters $\sigmaSUBu>0$ and $0\le\rhoSUBu\le1$ and with
$\CSUBK\equiv 2^{K}\pi^{(K-1)/2}\Gamma\big(\smhalf(K+1)\big)$.
Here $\bdelta_0$ denotes the $K$-variate Dirac delta function at $\bzero_K$, the $K\times1$ 
vector of zeroes. 

Kyung \textit{et al.} (2010) use the phrase \emph{group lasso} for the family of 
priors defined by (\ref{eq:DontArgue}) in the $\rhoSUBu=1$ special case. This naming
is due to the group LASSO methodology of Yuan \myand Lin (2006). The essence of
Yuan \myand Lin's (2006) extension of the ordinary LASSO is that particular
vectors coefficients, $\btheta$ say, are treated together as an entity and
penalty terms of the form $\lambda\Vert\btheta\Vert$, for some $\lambda>0$,
allow for all entries of $\btheta$ to be estimated as exactly zero.
In their frequentist approach to generalized additive model selection 
Chouldechova \myand Hastie (2015) apply this idea to vectors
of spline coefficients, denoted in this section by $\bu$. This allows for smooth 
function effects to be categorized as either linear or non-linear depending on whether 
$\buhat=\bzero$ or $\buhat\ne\bzero$, where $\buhat$ is an estimate of $\bu$.
In keeping with (\ref{eq:TwinCities}), we extend the group lasso distribution
to a $K$-variate ``spike-and-slab'' form. Note that (\ref{eq:DontArgue}) has
a point mass at $\bzero_K$, the $K$-vector of zeroes.

The scale parameter has the following prior distributions:
$$\sigmaSUBu\sim\mbox{Half-Cauchy}(\sSUBu)
$$
for hyperparameter $\sSUBu>0$. The prior distribution
justification given at the end of Section \ref{sec:modLinCoeff} also applies here.
The mixture parameter $\rhoSUBu$ is a user-specified hyperparameter.

\subsection{Hyperparameter Default Values}\label{sec:HYPdflt}

The standardization of the input data invokes scale invariance and justifies setting 
the hyperparameters to fixed constant values. With noninformativity in mind, 
our recommended default values of the hyperparameters are:
$$\sigma_{\beta_0}=10^5,\ \ \sSUBbeta=\sSUBeps=s_u=1000,\ \ \rhoSUBbeta=\rhoSUBu=\smhalf.$$
These values are used in the upcoming numerical studies and examples.

\subsection{Auxiliary Variable Representations}\label{sec:auxVarReps}

Distributional specifications such as (\ref{eq:TwinCities}) and (\ref{eq:DontArgue})
are not amenable to Markov chain Monte Carlo and mean field variational Bayes fitting algorithms
due to their non-standard full conditional distributions. In this subsection we re-express 
them using auxiliary variables, which are tailored so that all full conditional
distributions have standard forms. 

First, note that $\sigma\sim\mbox{Half-Cauchy}(s)$ is equivalent to
$$\sigma^2|a\sim\mbox{Inverse-Gamma}(\smhalf,1/a),\quad a\sim\mbox{Inverse-Gamma}(\smhalf,1/s).$$
For the case of (\ref{eq:TwinCities}) we introduce auxiliary variables 
$\gammaSUBbeta$, $\betatilde$ and $\bSUBbeta$ and re-define $\beta$ such 
that 
\begin{equation}
\beta=\gammaSUBbeta\betatilde,\ \ 
\gammaSUBbeta\sim\mbox{Bernoulli}(\rhoSUBbeta),\ \ 
\betatilde|\bSUBbeta,\sigsqSUBbeta\sim N(0,\sigmaSUBbeta^2/\bSUBbeta)
\ \ \mbox{and}\ \ \bSUBbeta\sim\mbox{Inverse-Gamma}(1,\smhalf).
\label{eq:BuggerLuggs}
\end{equation}
Then standard distributional manipulations can be used to show that
(\ref{eq:BuggerLuggs}) is equivalent to (\ref{eq:TwinCities}).
Similarly, with the introduction of the random variable $\gammaSUBu$,
(\ref{eq:DontArgue}) is equivalent to
$$\bu=\gammaSUBu\butilde,\ \ 
\gammaSUBu\sim\mbox{Bernoulli}(\rhoSUBu),\ \ 
\butilde|\bSUBu,\sigsqSUBu\sim N(0,\sigmaSUBu^2\bI/\bSUBu)
\ \ \mbox{and}\ \ \bSUBu\sim\mbox{Inverse-Gamma}
\left(\textstyle{\frac{K+1}{2}},\smhalf\right)
$$
courtesy of a result provided in Section 3.1 of Kyung \textit{et al.} (2010) for
the $\rhoSUBu=1$ case. 

\subsection{The Full Gaussian Response Model}

Consider, first, the case where Gaussianity of the $y_i$s is reasonably assumed.
Suppose that we apply the modelling structures of Sections \ref{sec:fjFuncs}--\ref{sec:splineCoeffs}
across each of $\dLin$ entries of the $\bxLini$ and $\dNon$ entries of $\bxNoni$.
Let $\bbeta$ be the $(\dLin+\dNon)\times1$ vector containing all of the linear term
coefficients and $\bu_1,\ldots,\bu_{\dNon}$ be the full set of spline coefficient vectors,
where $\bu_j$ has dimension $K_j\times1$. Also, apply the auxiliary variable representations of 
Section \ref{sec:auxVarReps}. The resultant full model is:
\begin{equation}
{\setlength\arraycolsep{0pt}
\begin{array}{rcl}
&&\by|\beta_0,\bgammaSUBbeta,\bbetatilde,\gamma_{u1},\ldots,\gamma_{u\dNon},
\butilde_1,\ldots,\butilde_{\dNon},\sigeps^2\sim \\[1ex]
&&\qquad\qquad N\left(\bone_n\beta_0+\bX(\bgammaSUBbeta\odot\bbetatilde)
+{\displaystyle\sum_{j=1}^{\dNon}}\bZ_j(\gammaSUBuj\butilde_j),\sigeps^2\bI_n\right),
\quad\beta_0\sim N(0,\sigma_{\beta_0}^2),
\\[4ex]
&&\sigeps^2|\aSUBeps\sim\mbox{Inverse-Gamma}(\smhalf,1/\aSUBeps),
\quad \aSUBeps\sim\mbox{Inverse-Gamma}(\smhalf,1/\sSUBeps^2),\\[1ex]
&&\gammaSUBbetaj\simind\mbox{Bernoulli}(\rhoSUBbeta),
\quad\betatilde_j|\sigmaSUBbeta^2,\bSUBbetaj\simind N(0,\sigmaSUBbeta^2/\bSUBbetaj),
,\quad 1\le j\le\dLin+\dNon,
\\[1ex]
&&\bSUBbetaj\simind\mbox{Inverse-Gamma}(1,\smhalf),\quad 1\le j\le\dLin+\dNon,
\quad\gammaSUBuj\simind\mbox{Bernoulli}(\rhoSUBu),\quad 1\le j\le\dNon,\\[1ex]
&&\butilde_j|\sigma_{uj}^2,\bSUBuj\simind N(\bzero,(\sigma_{uj}^2/\bSUBuj)\bI_{K_j}),
\quad
\bSUBuj\simind\mbox{Inverse-Gamma}\big(\smhalf(K_j+1),\frac{1}{2}\big),
\quad 1\le j\le\dNon,\\[1ex]
&&\sigma_{\beta}^2|\abeta\sim\mbox{Inverse-Gamma}(\smhalf,1/\abeta),
\quad \abeta\sim\mbox{Inverse-Gamma}(\smhalf,1/\sSUBbeta^2),\\[1ex]
&&\sigma_{uj}^2|a_{uj}\simind\mbox{Inverse-Gamma}(\smhalf,1/a_{uj}),
\quad a_{uj}\simind\mbox{Inverse-Gamma}(\smhalf,1/s_u^2),\quad 1\le j\le\dNon.\\[1ex]
\end{array}
}
\label{eq:mainModel}
\end{equation}
In (\ref{eq:mainModel}) we have $\bgammaSUBbeta\equiv[\gammaSUBbetaOne,\ldots,\gammaSUBbetaEnd]^T$.
Here, and elsewhere, the notation $\simind$ is an abbreviation 
for ``distributed independently as''.

\begin{figure}[!t]
\centering
\includegraphics[width=0.8\textwidth]{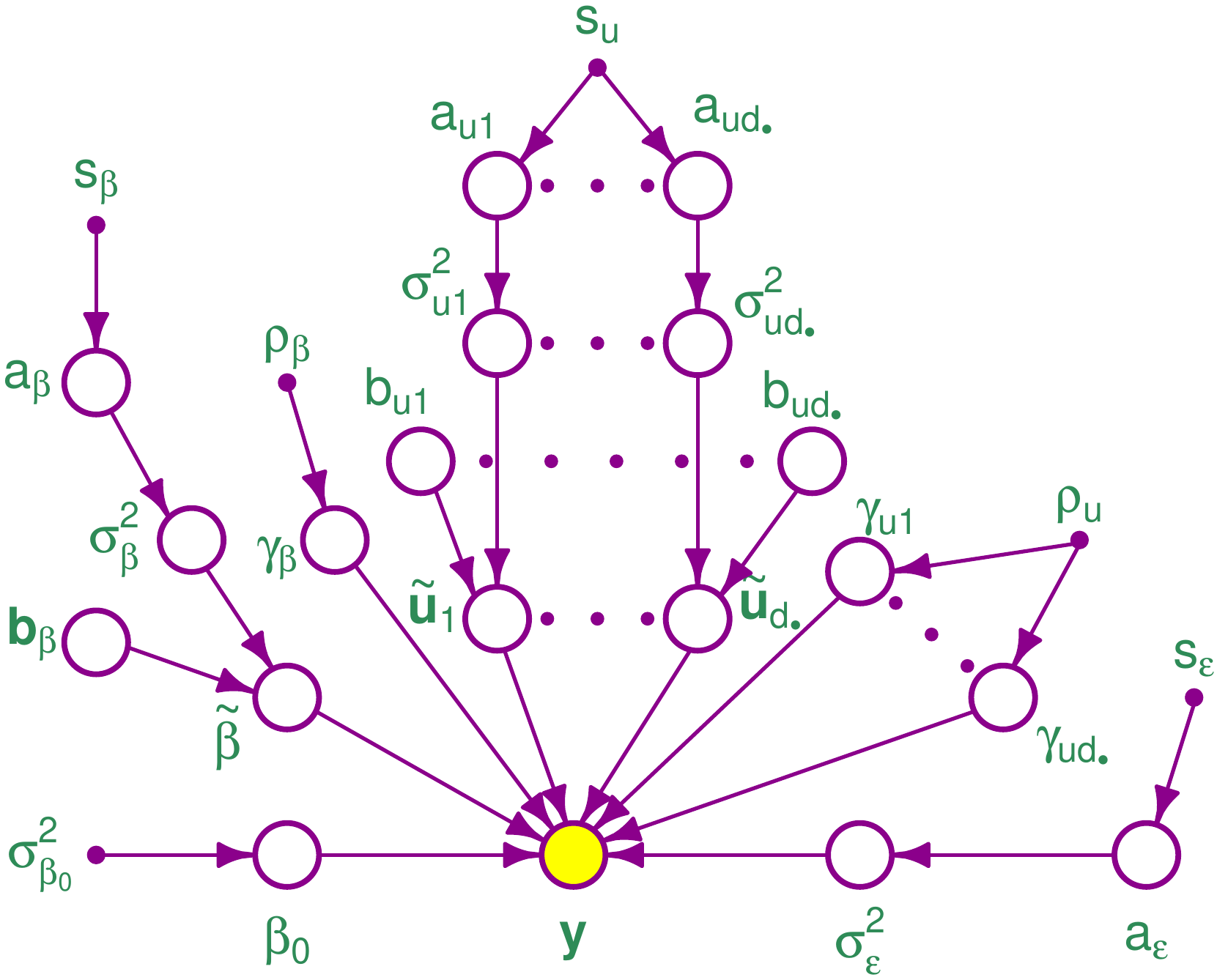}
\caption{
\textit{
Directed acyclic graph representation of Bayesian model (\ref{eq:mainModel}).
Random variables and vectors are shown as larger open circles, with shading indicating
to the observed response data. The small closed circles are user-specified hyperparameters.
}}
\label{fig:BayeGaussGAMselDAGcol}
\end{figure}

The full set of hyperparameters in (\ref{eq:mainModel}) is:
$$\sigma_{\beta_0},\sSUBbeta,\sSUBeps,s_u>0\quad\mbox{and}\quad 0\le\rhoSUBbeta,\rhoSUBu\le1.$$
Figure \ref{fig:BayeGaussGAMselDAGcol} shows the directed acyclic graph corresponding
to (\ref{eq:mainModel}).

\subsection{Adjustment for Binary Responses}\label{sec:adjBinResp}

\begin{figure}[!t]
\centering
\includegraphics[width=0.65\textwidth]{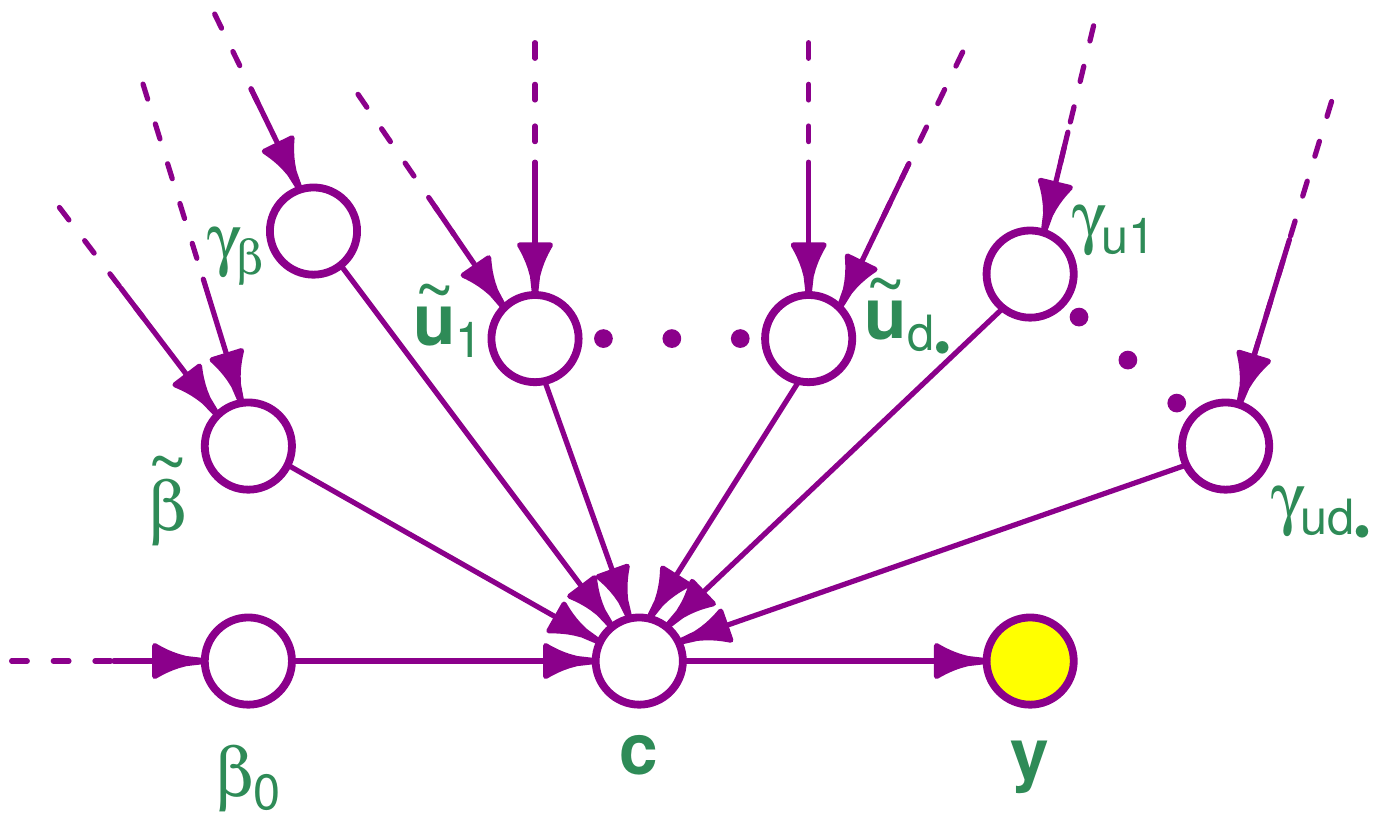}
\caption{\textit{
Sub-graph of the directed acyclic graph for the binary response adjustment
to (\ref{eq:mainModel}). This graph is the same as that shown in 
Figure \ref{fig:BayeGaussGAMselDAGcol}
except for locations near the response variables node. The new graph has the following
modifications: (1) the $\sigeps^2$ and $\aSUBeps$ nodes are absent, (2) a hidden node $\bc$ corresponding
to the Albert-Chib auxiliary variables is added to the position held by $\by$ in the Gaussian
response graph and the binary response observed data node $\by$ is a child of $\bc$. }}
\label{fig:BayeProbitGAMselSubDAG}
\end{figure}

Now suppose that the $y_i$ values are binary rather than continuous. 
Then an appropriate adjustment to (\ref{eq:mainModel}) is that where the 
likelihood is changed to 
\begin{equation}
{\setlength\arraycolsep{0pt}
\begin{array}{l}
y_i\big|\beta_0,\bgammaSUBbeta,\bbetatilde,\gamma_{u1},\ldots,\gamma_{u\dNon},
\butilde_1,\ldots,\butilde_{\dNon}\\[1ex]
\qquad\qquad\simind\mbox{Bernoulli}\Bigg(
\Phi\Bigg(\beta_0+\Big(\bX(\bgammaSUBbeta\odot\bbetatilde)
+{\displaystyle\sum_{j=1}^{\dNon}}\bZ_j(\gammaSUBuj\butilde_j)\Big)_i\Bigg)\Bigg).
\end{array}
}
\label{eq:MareeWatson}
\end{equation}
Following Albert \myand Chib (1993), we introduce auxiliary random variables 
$c_1,\ldots,c_n$ such that 
\begin{equation}
y_i=1\quad\mbox{if and only if}\quad c_i\ge 0
\label{eq:firstDonHearn}
\end{equation}
and impose the following conditional distribution on $\bc\equiv(c_1,\ldots,c_n)$:
\begin{equation}
{\setlength\arraycolsep{0pt}
\begin{array}{rcl}
&&\bc|\beta_0,\bgammaSUBbeta,\bbetatilde,\gamma_{u1},\ldots,\gamma_{u\dNon},
\butilde_1,\ldots,\butilde_{\dNon}\sim \\[1ex]
&&\qquad\qquad N\left(\bone_n\beta_0+\bX(\bgammaSUBbeta\odot\bbetatilde)
+{\displaystyle\sum_{j=1}^{\dNon}}\bZ_j(\gammaSUBuj\butilde_j),\bI_n\right).
\end{array}
}
\label{eq:ConeHead}
\end{equation}
The prior distributions on $\beta_0,\bgammaSUBbeta,\bbetatilde,\gamma_{u1},\ldots,\gamma_{u\dNon}$
and $\butilde_1,\ldots,\butilde_{\dNon}$ are the same as in the Gaussian response case. 
The error variance variables $\sigeps^2$ and $\aSUBeps$ are not present for binary responses. 
Therefore, our binary response model is a modification of (\ref{eq:mainModel}) for which
the $\by$ distributional specification is replaced by (\ref{eq:firstDonHearn}) and 
(\ref{eq:ConeHead}). Figure \ref{fig:BayeProbitGAMselSubDAG} shows this modification
in graphical terms.

\section{Practical Fitting and Model Selection}\label{sec:pracFitSel}

Practical generalized additive model selection based on the models 
described in Section \ref{sec:modelDescrip} requires approximation of
the posterior distributions of each of the hidden nodes (unshaded
circles) in Figures \ref{fig:BayeGaussGAMselDAGcol} and \ref{fig:BayeProbitGAMselSubDAG}.
The problem reduces to approximation conditional marginalization of 
directed acyclic graphs. The most accurate practical approach is 
Markov chain Monte Carlo (e.g. Gelfand \myand Smith, 1990). For the Gaussian response 
model (\ref{eq:mainModel}) and its binary response adjustment described in 
Section \ref{sec:adjBinResp}, Section \ref{sec:MCMC} provides full algorithmic details 
for Markov chain Monte Carlo-based approximate conditional marginalization. A faster, but 
less accurate, alternative is mean field variational Bayes (e.g. Wainwright \myand Jordan, 2008).
To facilitate scalability to very large data sets, we also provide a variational approximate
conditional marginalization algorithm in Section \ref{sec:MFVB}. Both approaches
have steps that depend on the data only through particular sufficient
statistic quantities. Therefore, there are considerable speed gains from 
computing and storing these quantities as part of a pre-preprocessing phase. 

\subsection{Pre-Processing and Storage of Key Matrices}\label{sec:preProc}

Algorithm \ref{alg:suffStats} is an important part of our overall strategy for fitting
our Bayesian generalized additive models in a stable and efficient manner. The first
steps involve standardizing the input data and storing the linear transformation
parameters to allow conversion of the final results to the original units.
Then design matrices denoted by $\bX$ and $\bZ$ are computed, with the latter containing
all required spline basis functions of the transformed predictor data. Lastly, sufficient 
statistic matrices such as $\bX^T\by$ and $\bZ^T\bZ$ are computed and stored --
ready for use in the upcoming Markov chain Monte Carlo and variational algorithms.

In Algorithm \ref{alg:suffStats} and the upcoming discussion and algorithms we use
the identifiers:
$$\XTy,\ \XTX,\ \XTy,\ \ZTy,\ \ZTX\quad \mbox{and}\quad \ZTZ$$
for storage of the sufficient statistic quantities $\bX^T\by$, $\bX^T\bX$,
$\bX^T\by$, $\bZ^T\by$, $\bZ^T\bX$ and $\bZ^T\bZ$.

\begin{algorithm}[!th]
\begin{center}
\begin{minipage}[t]{160mm}
\begin{small}
\begin{itemize}
\item[] Inputs: $\byOrig (n\times1)$;\quad
$\bxLinOrigj (n\times1)$,\ $1\le j\le\dLin$;\quad $\bxNonOrigj (n\times1)$,\ $1\le j\le\dNon$
\item[]  $\meanyOrig\thickarrow\mbox{sample mean of $\byOrig$}\ \ ;\ \ 
\stdevyOrig\thickarrow\mbox{sample standard dev'n of $\byOrig$}$
\item[] If $\byOrig$ is continuous then $\by\thickarrow\big\{\byOrig-\meanyOrig\bone_n\big\}/\stdevyOrig$
\item[] If $\byOrig$ is binary then $\by\thickarrow\byOrig$
\item[] For $j=1,\ldots,\dLin$:
\begin{itemize}
\item[] $\meanxOrigLinj\thickarrow\mbox{sample mean of $\bxLinOrigj$}\ \ ;\ \ 
\stdevxOrigLinj\thickarrow\mbox{sample standard dev'n of $\bxLinOrigj$}$
\item[] $\bxLinj\thickarrow\big\{\bxLinOrigj-\meanxOrigLinj\bone_n\big\}/\stdevxOrigLinj$
\end{itemize}
\item[] For $j=1,\ldots,\dNon$:
\begin{itemize}
\item[] $\meanxOrigNonj\thickarrow\mbox{sample mean of $\bxNonOrigj$}\ \ ;\ \ 
\stdevxOrigNonj\thickarrow\mbox{sample standard dev'n of $\bxNonOrigj$}$
\item[] $\bxNonj\thickarrow\big\{\bxNonOrigj-\meanxOrigNonj\bone_n\big\}/\stdevxOrigNonj$
\end{itemize}
\item[] $\bX\thickarrow\Big[\bxLinOne\cdots\bxLindLin\ \bxNonOne\cdots\bxNondNon\Big]$
\item[] For $j=1,\ldots,\dNon$:
\begin{itemize}
\item[]  $\bZ_j\thickarrow\textrm{$n\times K_j$ matrix containing the canonical 
Demmler-Reinsch basis for the predictor}$
\item[] $\qquad\quad\ \ \textrm{data vector $\bxNonj$, using the construction 
described in Section \ref{sec:ZcDR} of the supplement}$
\end{itemize}
\item[] $\bZ\thickarrow\big[\bZ_1\ \cdots\ \bZ_{\dNon}\big]$\ \ ;\ \ $\XTy\thickarrow\bX^T\by$
\ \ ;\ \ $\XTX\thickarrow\bX^T\bX$\ \ ;\ \  $\ZTy\thickarrow\bZ^T\by$\\[-3ex]
\item[] $\ZTX\thickarrow\bZ^T\bX$\ \ ;\ \ $\ZTZ\thickarrow\bZ^T\bZ$
\item[] Outputs: $\by$,\ $\bX$,\ $\bZ_1,\ldots,\bZ_{\dNon}$,   
$\XTy$,\ $\XTX$,\ $\ZTy$,\ $\ZTX$,\ $\ZTZ$,\ $\meanyOrig$,\ $\stdevyOrig$, \\[-3ex]
\item[] $\qquad\qquad\
\Big\{\big(\meanxOrigLinj,\stdevxOrigLinj\big):1\le j\le\dLin\Big\}$,\ 
$\Big\{\big(\meanxOrigNonj,\stdevxOrigNonj\big):1\le j\le\dNon\Big\}$
\end{itemize}
\end{small}
\end{minipage}
\end{center}
\caption{\textit{Pre-processing of original data and creation of key matrices for
input into Bayesian generalized additive model algorithms.}}
\label{alg:suffStats}
\end{algorithm}

\subsection{Notation Used in the Fitting Algorithms}

For the main fitting algorithms it is useful to have the following definitions in place:
\begin{equation}
\begin{tabular}{l}
\mbox{$K_j\equiv\mbox{the number of columns in $\bZ_j$},\quad 1\le j\le \dNon$,}\\[1ex]
\mbox{$\gothicc$ is the $(\dNon+1)\times1$ vector with entries 
$\gothicc_1\equiv0$ and $\gothicc_{j+1}\equiv\sum_{k=1}^j K_k,\ 1\le j\le \dNon$,}\\[1ex]
\mbox{$\ZTyj\equiv\mbox{the sub-block of $\ZTy$ 
corresponding to rows $(\gothicc_j+1)$ to $\gothicc_{j+1}$},\ 1\le j\le \dNon$,}\\[1ex]
\mbox{$\ZTXj\equiv\mbox{the sub-block of $\ZTX$ 
corresponding to rows $(\gothicc_j+1)$ to $\gothicc_{j+1}$},\ 1\le j\le\dNon$,}\\[1ex]
\mbox{$\ZTZjjd\equiv\mbox{the sub-block of $\ZTZ$ 
corresponding to rows $(\gothicc_j+1)$ to $\gothicc_{j+1}$}$}\\[1ex]
\mbox{$\qquad\qquad\quad\ \mbox{and columns $(\gothicc_{j'}+1)$ to $\gothicc_{j'+1}$},\ 1\le j,j'\le\dNon.$}
\end{tabular}
\label{eq:PaulBugden}
\end{equation}
Note that, according to the notation in (\ref{eq:PaulBugden}), 
$$\ZTXj=\bZ_j^T\bX\quad\mbox{and}\quad\ZTZjjd=\bZ_j^T\bZ_{j'}.$$
The updates in approximate inference iterative algorithms, presented in Sections \ref{sec:MCMC} 
and \ref{sec:MFVB}, depend on particular columns and rows of the matrices listed in
(\ref{eq:PaulBugden}). These will be specified using the following notational convention:
$\be_r$ is a column vector of appropriate length with $r$th entry equal to $1$ and zeroes elsewhere.
For example, the $j$th column of $\XTX$ is $\XTX\be_j$ where $\be_j$ is the $(\dLin+\dNon)\times1$ 
vector with $j$th entry 1 and $0$ elsewhere.
Implementations of the upcoming algorithms normally would not require explicit
calculation and storage of $\be_r$ vectors and, instead, array subsetting code 
specific to the programming language can be used. However, for algorithm listing 
use of the $\be_r$ notation has the advantage of avoiding further subscripting.

To allow the Gaussian and Bernoulli response cases to be handled together we also
use the notation $\yToneadj$, $\XTyadj$ and $\ZTyadj$. These are adjustments
of $\yTone$, $\XTy$ and $\ZTy$ in which the $\by$ vector is replaced by $\bc$:
the Albert-Chib auxiliary variables vector that arises in the Bernoulli response
case. The notation of (\ref{eq:PaulBugden}) for extraction of sub-blocks of $\ZTy$
also applies to $\ZTyadj$.

The main algorithms also uses the following functions:
$$\logit(x)\equiv \log\left(\frac{x}{1-x}\right),\ \ 
\expit(x)\equiv\logit^{-1}(x)=\frac{1}{1+\exp(-x)}
\ \ \mbox{and}\ \ \zeta(x)=\log\{2\Phi(x)\}
$$
where, as before, $\Phi$ is the $N(0,1)$ cumulative distribution function.
It follows that $\zeta'(x)=\phi(x)/\Phi(x)$, where $\phi$ is the $N(0,1)$ density function, 
which arises in Algorithm \ref{alg:MFVB}. Stable computation of $\zeta'(x)$ 
when $x$ is a large negative number is not straightforward. Azzalini (2023) 
and Wand \myand Ormerod (2012), for example, provide practical solutions
to this problem. Lastly, an expression of the form $\zeta'(\bv)$, where $\bv$
is a column vector, is such that function evaluation is element-wise.

\subsection{Markov Chain Monte Carlo}\label{sec:MCMC}

For the Bayesian graphical model (\ref{eq:mainModel}) and the binary response
adjustments given in Section \ref{sec:adjBinResp}, determination of each of the 
full conditional distributions for Markov Chain Monte Carlo sampling is fairly
straightforward. Virtually all of the full conditional distributions have standard forms
such as Bernoulli, Beta, Inverse Gamma  and Multivariate Normal distributions. 
Possible exceptions are the Inverse Gaussian and Truncated Normal distributions,
but are such that effective solutions are provided, respectively, by Michael 
\textit{et al.} (1976) and Robert (1995). Therefore, Markov Chain Monte Carlo 
sampling essentially reduces to Gibbs sampling for the models at hand.
Algorithm \ref{alg:MCMC} lists the full set of steps needed to draw 
samples from the posterior distributions of the model parameters. The fact that
most of the draws only require the sufficient statistic matrices from
Algorithm \ref{alg:suffStats} means that the sampling can be done quite
rapidly regardless of sample size.

\begin{algorithm}[!th]
\begin{center}
\begin{minipage}[t]{160mm}
\begin{small}
\begin{itemize}
\item[] Data Inputs: $\by\ (n\times1)$;
\ \ $\bX\ \big(n\times(\dLin+\dNon)\big)$;
\ \ $\bZ_j\ (n\times K_j)$, $1\le j\le\dNon$.
\item[] Response Type Input: $\textbf{responseType}\in\{\mbox{Gaussian},\mbox{Bernoulli}\}.$
\item[] Sufficient Statistics Inputs: $\XTy$,\ $\XTX$,\ $\ZTy$,\ $\ZTX$, $\ZTZ$.
\item[] Hyperparameter Inputs: $\sigma_{\beta_0},\sSUBbeta,\sSUBeps,s_u>0,\ \ 0\le\rhoSUBbeta,\rhoSUBu\le1$.
\item[] Chain Length Inputs: $\Nwarm$ and $\Nkept$, both positive integers.\\[-3ex]
\item[] Initialize: $\bgammaSUBbeta^{[0]}\thickarrow\smhalf\bone_{\dLin+\dNon}$; 
$\gammaSUBuj^{[0]}\thickarrow\smhalf$, $1\le j\le\dNon$;\ \ 
$\bbetatilde^{[0]}\thickarrow\bzero_{\dLin+\dNon}$\\[-3ex]
\item[]$\qquad\qquad\butilde_j^{[0]}\thickarrow\bzero_{K_j}$, $1\le j\le\dNon$;\ \ 
$\aSUBeps^{[0]}\thickarrow 1$;\ \ 
$(\sigeps^2)^{[0]}\thickarrow 1$;\ \ $(\sigmaSUBbeta^2)^{[0]}\thickarrow 1$\ ;
\ $a_{\beta}^{[0]}\thickarrow 1$\\[-3ex]
\item[]$\qquad\qquad\bbSUBbeta^{[0]}\thickarrow \bone_{\dLin+\dNon}$\ \ ;\ \   
$\bSUBuj^{[0]}\thickarrow 1$,\ $1\le j\le\dNon$\\[-3ex]
\item[]$\qquad\qquad a_{uj}^{[0]}\thickarrow 1$,\ $1\le j\le\dNon$;\ \  
$(\sigma_{uj}^2)^{[0]}\thickarrow 1$,\ $1\le j\le\dNon$.\\[-3ex]
\item[] $\yToneadj\thickarrow 0$\ \ ;\ \ $\XTyadj\thickarrow\XTy$
\ \ ;\ \ $\ZTyadj\thickarrow\ZTy$\\[-3ex]
\item[] For $j=1,\ldots,\dNon$:$\quad\bwZj\thickarrow\mbox{diagonal}\big(\ZTZjj\big)$
\item[] For $g=1,\ldots,\Nwarm+\Nkept$:\\[-3ex]
\begin{itemize}
\item[] $\omegaAAA\thickarrow \yToneadj$\\[-2ex]
\item[] $\omegaBBB\thickarrow\big\{n\big/(\sigeps^2)^{[g-1]}\big\} + (1/\sigma_{\beta_0}^2)$
\ \ \ ;\ \ \ 
$\beta_0^{[g]}\sim
\displaystyle{N\left(\frac{\omegaAAA}{(\sigeps^2)^{[g-1]}\omegaBBB},\frac{1}{\omegaBBB}\right)}$\\[-1ex]
\item[] $\bOmega\thickarrow \left(\bgammaSUBbeta^{[g-1]}\bgammaSUBbeta^{[g-1]T}\right)\odot(\XTX)\Big/
(\sigeps^2)^{[g-1]}+\diag\left(\bbSUBbeta^{[g-1]}\right)\Big/(\sigmaSUBbeta^2)^{[g-1]}$\\[-1ex]
\item[] $\omegaCCC\thickarrow \XTyadj
-{\displaystyle\sum_{j=1}^{\dNon}}\ZTXjT
\big(\gammaSUBuj^{[g-1]}\butilde_j^{[g-1]}\big)$\\[-1.5ex]
\item[] Decompose $\bOmega=\bUOmega\diag(\bdOmega)\bUOmega^T$ where $\bUOmega\bUOmega^T=\bI$\\[-3ex]
\item[]$\bz\sim N(\bzero,\bI)\ \big((\dLin+\dNon)\times1\big)$\ \ \ ;\ \ \  
$\bbetatilde^{[g]}\thickarrow 
\bUOmega\displaystyle{\left(\frac{\bUOmega^T\bz}{\sqrt{\bdOmega}}
+\frac{\bUOmega^T\left(\bgammaSUBbeta^{[g-1]}\odot\omegaCCC\right)}
{\bdOmega(\sigeps^2)^{[g-1]}}\right)}$\\[-1ex]
\item[] $\big(\bbSUBbeta^{[g]})_j\sim\mbox{Inverse-Gaussian}
\Big(\sigmaSUBbeta^{[g-1]}\Big/
\Big|\big(\bbetatilde^{[g]}\big)_j\Big|\Big),\quad 1\le j\le \dLin+\dNon$\\[-2ex]
\item[] $(\sigmaSUBbeta^2)^{[g]}\sim
\mbox{Inverse-Gamma}\displaystyle{\left(\smhalf(\dLin+\dNon+1),1/a_{\beta}^{[g-1]}
+\smhalf\bbetatilde^{[g]T}\diag\left(\bbSUBbeta^{[g]}\right)\bbetatilde^{[g]}\right)}$
\item[] $\abeta^{[g]}\sim\mbox{Inverse-Gamma}
\displaystyle{\left(1,\big\{1\big/(\sigmaSUBbeta^2)^{[g]}\big\}+(1/\sSUBbeta^2)\right)}$
\ \ ;\ \ 
$\bbetaCurr\thickarrow \bgammaSUBbeta^{[g-1]}\odot\bbetatilde^{[g]}$
\item[] For $j=1,\ldots,\dNon$: 
$\quad\bujCurr\thickarrow \gammaSUBuj^{[g-1]}\butilde_j^{[g-1]}$
\item[] For $j=1,\ldots,\dLin+\dNon$:\\[-3ex]
\begin{itemize}
\item[] 
$\omegaDDD\thickarrow \be_j^T\XTyadj-\big(\XTX\be_j\big)_{-j}^T\big(\bbetaCurr)_{-j}
-\displaystyle{\sum_{j'=1}^{\dNon}}\Big(\ZTXjd\be_j\Big)^T\bujdCurr$
\item[] $\omegaEEE\thickarrow\logit(\rhoSUBbeta)-
\smhalf\Big\{\big(\betatilde_j^{[g]}\big)^2\be_j^T\XTX\be_j
-2\betatilde_j^{[g]}\omegaDDD\Big\}\Big/(\sigeps^2)^{[g-1]}$
\item[] $\big(\bgammaSUBbeta^{[g]})_j\sim\mbox{Bernoulli}\big(\expit(\omegaEEE)\big)$
\end{itemize}
\item[] \textit{continued on a subsequent page $\ldots$}\\[-2ex]
\end{itemize}
\end{itemize}
\end{small}
\end{minipage}
\end{center}
\caption{\textit{Markov chain Monte Carlo generation of samples from the posterior 
distributions of the parameters in (\ref{eq:mainModel}).}}
\label{alg:MCMC}
\end{algorithm}

\setcounter{algorithm}{1}
\begin{algorithm}[!th]
\begin{center}
\begin{minipage}[t]{160mm}
\begin{small}
\begin{itemize}
\item[]
\begin{itemize}
\item[] $\bbetaCurr\thickarrow \bgammaSUBbeta^{[g]}\odot\bbetatilde^{[g]}$\ \ \ ;\ \ \ 
For $j=1,\ldots,\dNon$: $\quad\butildejCurr\thickarrow\butilde_{j}^{[g-1]}$
\item[] For $j=1,\ldots,\dNon$:\\[-2ex]
\begin{itemize}
\item[]$\omegaFFF\thickarrow\ZTyjadj-\ZTXj\bbetaCurr
-{\displaystyle\sum_{j'\ne j}^{\dNon}}\ZTZjjd\big(\gammaSUBujd^{[g-1]}\butildejdCurr\big)$ 
\item[] 
$\omegaGGG\thickarrow 
\Big\{\gammaSUBuj^{[g-1]}\bwZj\Big/(\sigeps^2)^{[g-1]}\Big\}
+\Big\{\bSUBuj^{[g-1]}\bone_{K_j}\Big/(\sigma_{uj}^2)^{[g-1]}\Big\}$
\item[]$\bz\sim N(\bzero,\bI_{K_j})$\ \ \ ;\ \ \ 
$\butildejCurr\thickarrow \big(\bz\big/\sqrt{\omegaGGG}\big)
+\big[\gammaSUBuj^{[g-1]}\omegaFFF\big/\big\{\omegaGGG(\sigeps^2)^{[g-1]}\big\}\big]$
\end{itemize}
\item[] For $j=1,\ldots,\dNon$: $\quad\butilde_{j}^{[g]}\thickarrow\butildejCurr$
\item[] For $j=1,\ldots,\dNon$:
\begin{itemize}
\item[] $\bSUBuj^{[g]}\sim\mbox{Inverse-Gaussian}
\left(\sigma_{uj}^{[g-1]}\Big/\Vert\butilde_j^{[g]}\Vert,1\right)$
\item[] $(\sigma^2_{uj})^{[g]}\sim\mbox{Inverse-Gamma}\left(\smhalf(K_j+1),\big\{1\big/a_{uj}^{[g-1]}\big\}
+\smhalf\Vert\butilde_j^{[g]}\Vert^2\bSUBuj^{[g]}\right)$
\item[] $a_{uj}^{[g]}\sim\mbox{Inverse-Gamma}
\displaystyle{\left(1,\{1\big/(\sigma^2_{uj})^{[g]}\}+(1/s_u^2)\right)}$
\end{itemize}
\item[] 
For $j=1,\ldots,\dNon$: $\quad\gammaSUBujCurr\thickarrow\gammaSUBuj^{[g-1]}$\\[-3ex]
\item[]For $j=1,\ldots,\dNon$:\\[-4ex]
\begin{itemize}
\item[]$\omegaHHH\thickarrow\ZTyjadj-\ZTXj\bbetaCurr
-\displaystyle{\sum_{j'\ne j}^{\dNon}}\ZTZjjd\big(\gammaSUBujdCurr\,\butilde_{j'}^{[g]}\big)$\\[0ex]
\item[]$\omegaIII\thickarrow\logit\left(\rhoSUBu\right)-
\smhalf\Big\{\bwZj^T\big(\butilde_{j}^{[g]}\odot\butilde_{j}^{[g]}\big)
-2\big(\butilde_{j}^{[g]}\big)^T\omegaHHH\Big\}\Big/(\sigeps^2)^{[g-1]}$
\item[]$\gammaSUBujCurr\sim\mbox{Bernoulli}\big(\expit(\omegaIII)\big)$
\end{itemize}
\item[] 
For $j=1,\ldots,\dNon$: $\quad\gammaSUBuj^{[g]}\thickarrow\gammaSUBujCurr$\\[-2ex]
\item[] $\omegaJJJ\thickarrow \bone_n\beta_0^{[g]}+
\bX\left(\bgammaSUBbeta^{[g]}\odot\bbetatilde^{[g]}\right)
+\displaystyle{\sum_{j=1}^{\dNon}}\bZ_j\left(\gammaSUBuj^{[g]}\,\butilde_j^{[g]}\right)$
\item[] If $\textbf{responseType}\ \mbox{is Gaussian}$ then
\begin{itemize}
\item[] $(\sigeps^2)^{[g]}\sim\mbox{Inverse-Gamma}
\left(\smhalf(n+1),\big(1\big/\aSUBeps^{[g-1]}\big)+\smhalf\Vert\by-\omegaJJJ\Vert^2\right)$ 
\item[] $\aSUBeps^{[g]}\sim\mbox{Inverse-Gamma}
\displaystyle{\left(1,\big\{1\big/(\sigeps^2)^{[g]}\big\}+(1/\sSUBeps^2)\right)}$
\end{itemize}
\item[] If $\textbf{responseType}\ \mbox{is Bernoulli}$ then
\begin{itemize}
\item[] $(\sigeps^2)^{[g]}\thickarrow 1$
\item[] For $i=1,\ldots,n$: 
\begin{itemize}
\item[] $\omegaKKK\sim\mbox{Truncated-Normal}_+\big((2y_i-1)(\omegaJJJ)_i,1\big)$\ \ ;\ \ 
$c_i\thickarrow (2y_i-1)\omegaKKK$
\end{itemize}
\item[] $\yToneadj\thickarrow \bone^T\bc$\ \ ;\ \  $\XTyadj\thickarrow \bX^T\bc$
\ \ ;\ \  $\ZTyadj\thickarrow \bZ^T\bc$
\end{itemize}
\end{itemize}
\item[] Outputs: All chains after omission of the first $\Nwarm$ values.
\end{itemize}
\end{small}
\end{minipage}
\end{center}
\caption{\textbf{continued.}\ \textit{This is a continuation of the description of this algorithm that
commences on a preceding page.}}
\end{algorithm}

\subsection{Mean Field Variational Bayes}\label{sec:MFVB}

Mean field variational Bayes approximate fitting and inference for (\ref{eq:mainModel}) involves
approximation of the joint posterior density function of the model parameters by a product density 
form such as
\begin{equation}
{\setlength\arraycolsep{0pt}
\begin{array}{rcl}
&&\pDens\big(\beta_0,\bgammaSUBbeta,\bbetatilde,\bgammaSUBu,
\butilde,\bbSUBbeta,\sigsqSUBbeta,\aSUBbeta,
\bbSUBu,\bsigsqSUBu,\baSUBu,\sigeps^2,\aSUBeps|\by\big)\\[1ex]
&&\quad\qquad\approx\qDens(\beta_0)\qDens(\bgammaSUBbeta)\qDens(\bbetatilde)\qDens(\bgammaSUBu)
\qDens(\butilde)\qDens(\bbSUBbeta)\qDens(\sigma_{\beta}^2)\qDens(\aSUBbeta)
\qDens(\bbSUBu)\qDens(\bsigsqSUBu)\qDens(\baSUBu)\qDens(\sigeps^2)
\qDens(\aSUBeps)
\end{array}
}
\label{eq:PrincessMartha}
\end{equation}
where, for example, $\butilde\equiv(\butilde_1,\ldots,\butilde_{\dNon})$  
and $\bbSUBbeta\equiv\big(\bSUBbetaOne,\ldots,\bSUBbetaLast\big)$. There are
numerous options for the stringency of the product restriction and 
the choice involves trade-offs concerning tractability, accuracy and speed.
For example, one could contemplate replacing $\qDens(\beta_0)\qDens(\bbetatilde)\qDens(\butilde)$
in (\ref{eq:PrincessMartha}) by $\qDens(\beta_0,\bbetatilde,\butilde)$ and 
improve the accuracy of approximation. However, the more stringent approximation
is less tractable. In addition to the product restriction (\ref{eq:PrincessMartha})
we also impose the product density restrictions:
\begin{equation}
\qDens(\gammaSUBbeta)=\prod_{j=1}^{\dLin+\dNon}\qDens(\gammaSUBbetaj)
,\quad\qDens(\butilde)=\prod_{j=1}^{\dNon}\qDens(\butilde_j)
\quad\mbox{and}\quad
\qDens(\bgammaSUBu)=\prod_{j=1}^{\dNon}\qDens(\gammaSUBuj).
\label{eq:MarkSuters}
\end{equation}

With the product density restrictions in place, we obtain the optimal $\qDens$-densities
by minimising the Kullback-Leibler divergence of the left-hand side of (\ref{eq:PrincessMartha})
from the right-hand side. The optimal $\qDens$-density forms can be expressed in
terms of the full conditional density functions as given by equation (6) of Ormerod \myand Wand (2010).
The optimal $\qDens$-density parameters can then be solved via a coordinate ascent algorithm.
Since each of the full conditionals in the models at hand have standard forms, the optimal
$\qDens$-density functions are relatively simple and the coordinate ascent updates have
closed forms.

The Bayesian graphical model for wavelet regression described in  Section 3 of Wand \myand Ormerod (2011) 
is similar in nature to the generalized additive selection model (\ref{eq:mainModel}). 
Hence, the relevant details on the requisite mean field variational Bayes calculations 
for (\ref{eq:mainModel}) can be gleaned from the $\qDens$-density derivations given in 
Appendix D of Wand \myand Ormerod (2011).

Some examples of the resulting optimal $\qDens$-density forms are:
\begin{equation}
\begin{array}{rcl}
&&\qDens(\bbetatilde)\ \mbox{has a}\ N\big(\bmu_{\qDens(\bbetatilde)},
\bSigma_{\qDens(\bbetatilde)}\big)\ \mbox{density function, and }\\[0ex]
&&\qDens(\sigsqSUBeps)\ \mbox{has an}\ \mbox{Inverse-Gamma}
\big(\kappa_{\qDens(\sigsqSUBeps)},\lambda_{\qDens(\sigsqSUBeps)}\big)
\ \mbox{density function.}
\end{array}
\label{eq:pumpkinCarve}
\end{equation}
The optimal Inverse Gamma shape parameter $\kappa_{\qDens(\sigsqSUBeps)}$ has explicit 
solution $\smhalf(n+1)$. However, the equations for the 
optimal values of $\bmu_{\qDens(\bbetatilde)}$, $\bSigma_{\qDens(\bbetatilde)}$ 
and $\lambda_{\qDens(\sigsqSUBeps)}$ are interdependent and iteration is 
required to obtain their optimal values. Algorithm \ref{alg:MFVB} lists the
full set of steps required to obtain all $\qDens$-density parameters, with
notation similar to that used in (\ref{eq:pumpkinCarve}) for the other
$\qDens$-density parameters.

A final aspect of Algorithm \ref{alg:MFVB} is determination of good
stopping criteria for the coordinate ascent scheme. As is common
in the mean field variational Bayes literature we monitor relative
increases in the approximate marginal log-likelihood,  also known as the 
evidence lower bound, which we denote by $\log\punder(\by;\qDens)$.
Section \ref{sec:logML} of the supplement contains an explicit expression for 
the approximate marginal log-likelihood for the Section \ref{sec:modelDescrip} 
models under product restrictions (\ref{eq:PrincessMartha})--(\ref{eq:MarkSuters}).

\begin{algorithm}[!th]
\begin{center}
\begin{minipage}[t]{160mm}
\begin{small}
\begin{itemize}
\item[] Data Inputs: $\by\ (n\times1)$;
\ \ $\bX\ \big(n\times(\dLin+\dNon)\big)$;
\ \ $\bZ_j\ (n\times K_j)$, $1\le j\le\dNon$.
\item[] Response Type Input: $\textbf{responseType}\in\{\mbox{Gaussian},\mbox{Bernoulli}\}$.
\item[] Sufficient Statistics Inputs: $\XTy$,\ $\XTX$,\ $\ZTy$,\ $\ZTX$, $\ZTZ$
\item[] Hyperparameter Inputs: $\sigma_{\beta_0},\sSUBbeta,\sSUBeps,s_u>0,\ \ 0\le\rhoSUBbeta,\rhoSUBu\le1$.
\item[] Convergence Criterion Input: $\varepsilon_{\tiny\mbox{toler.}}:\   
\mbox{a small positive number such as}\ 10^{-8}$.
\item[] Initialize: $\bmu_{\qDens(\gammaSUBbeta)}\thickarrow\smhalf\bone_{\dLin+\dNon}$
\ ;\ $\bmu_{\qDens(\bbetatilde)}\thickarrow\bzero_{\dLin+\dNon}$
\ ;\ $\mu_{\qDens(1/\aSUBeps)}\thickarrow1$,
\ ;\ $\mu_{\qDens(1/\sigeps^2)}\thickarrow1$
\\[-3ex]
\item[]$\qquad\qquad\ \mu_{\qDens(1/a_{\beta})}\thickarrow1$\ ;\  
        $\mu_{\qDens(1/\sigmaSUBbeta^2)}\thickarrow1$\ ;\ 
        $\kappa_{\qDens(\sigmaSUBbeta^2)}\thickarrow \smhalf(\dLin+\dNon+1)$\ ;\ 
        $\kappa_{\qDens(\abeta)}\thickarrow1$ 
\\[-3ex]
\item[]$\qquad\qquad\ \kappa_{\qDens(\sigeps^2)}\thickarrow\smhalf(n+1)\ ;\  
\kappa_{\qDens(\aSUBeps)}\thickarrow 1$\ ;\ 
$\bmu_{\qDens(\bbSUBbeta)}\thickarrow\bone_{\dLin+\dNon}$
\\[-2.5ex]
\item[] $\qquad\qquad\ \yToneadj\thickarrow 0$\ \ ;\ \ 
$\XTyadj\thickarrow\XTy$\ \ ;\ \ $\ZTyadj\thickarrow\ZTy$\\[-3ex]
\item[] $\qquad\qquad\ $For $j=1,\ldots,\dNon$:\\[-4ex]
\begin{itemize}
\item[]$\qquad\qquad\ \bmu_{\qDens(\butilde_j)}\thickarrow\bzero_{K_j}$\ ;\ 
$\bsigma^2_{\qDens(\butilde_j)}\thickarrow\bone_{K_j}$\ ; \
$\mu_{\qDens(\gammaSUBuj)}\thickarrow\smhalf$\\[-3ex]
\item[]$\qquad\qquad\ \mu_{\qDens(1/a_{uj})}\thickarrow1$\ ;\ 
       $\mu_{\qDens(1/\sigma_{uj}^2)}\thickarrow1$\ ;\ 
       $\mu_{\qDens(\bSUBuj)}\thickarrow1$\\[-3ex]
\item[] $\qquad\qquad\ \kappa_{\qDens(\sigma_{uj}^2)}\thickarrow \smhalf(K_j+1)$\ ;\  
$\kappa_{\qDens(a_{uj})}\thickarrow 1$\ ;\ 
$\bwZj\thickarrow\mbox{diagonal}\big(\ZTZjj\big)$
\end{itemize}
\item[] Cycle:\\[-4ex]
\begin{itemize}
\item[] $\omegaLLL\thickarrow \yToneadj$\\[-2ex]
\item[] $\sigma^2_{\qDens(\beta_0)}\thickarrow 
1\big/\{n\mu_{\qDens(1/\sigeps^2)}+(1/\sigma_{\beta_0}^2)\}$\ \ \ ;\ \ \  
$\mu_{\qDens(\beta_0)}\thickarrow\sigma^2_{\qDens(\beta_0)}\mu_{\qDens(1/\sigeps^2)}\omegaLLL$
\item[] $\bOmega_{\qDens(\gammaSUBbeta)}\thickarrow
\diag\big\{\bmu_{\qDens(\gammaSUBbeta)}\odot(\bone-\bmu_{\qDens(\gammaSUBbeta)})\big\}
+\bmu_{\qDens(\gammaSUBbeta)}\bmu_{\qDens(\gammaSUBbeta)}^T$
\item[] $\bSigma_{\qDens(\bbetatilde)}\thickarrow
\left\{\mu_{\qDens(1/\sigeps^2)}\bOmega_{\qDens(\gammaSUBbeta)}\odot
(\bX^T\bX)+\mu_{\qDens(1/\sigmaSUBbeta^2)}\diag(\bmu_{\qDens(\bbSUBbeta)})
\right\}^{-1}$
\item[] $\omegaMMM\thickarrow \XTyadj
-{\displaystyle\sum_{j=1}^{\dNon}}\ZTXjT
\big(\mu_{\qDens(\gammaSUBuj)}\,\bmu_{\qDens(\butilde_j)}\big)$\\[-1.5ex]
\item[] $\bmu_{\qDens(\bbetatilde)}\thickarrow\mu_{\qDens(1/\sigeps^2)}\bSigma_{\qDens(\bbetatilde)}
\big(\bmu_{\qDens(\gammaSUBbeta)}\odot\omegaMMM\big)$
\item[] $\omegaNNN\thickarrow\bmu_{\qDens(\bbetatilde)}\odot\bmu_{\qDens(\bbetatilde)}+
\mbox{diagonal}\big(\bSigma_{\qDens(\bbetatilde)}\big)$\ \ \ ;\ \ \ 
$\bmu_{\qDens(\bbSUBbeta)}\thickarrow \Big(\mu_{\qDens(1/\sigma_{\beta}^2)}\omegaNNN\Big)^{-1/2}$
\item[] $\lambda_{\qDens(\sigmaSUBbeta^2)}\thickarrow\mu_{\qDens(1/a_{\beta})}+
\smhalf\bmu_{\qDens(\bbSUBbeta)}^T\omegaNNN$\ \ \ ;\ \ \ 
$\mu_{\qDens(1/\sigmaSUBbeta^2)}\thickarrow\kappa_{\qDens(\sigmaSUBbeta^2)}/\lambda_{\qDens(\sigmaSUBbeta^2)}$
\item[] $\lambda_{\qDens(\abeta)}\thickarrow \mu_{\qDens(1/\sigmaSUBbeta^2)}+\sSUBbeta^{-2}$\ \ \ ;\ \ \ 
$\mu_{\qDens(1/\abeta)}\thickarrow \kappa_{\qDens(\abeta)}\big/\lambda_{\qDens(\abeta)}$
\item[] For $j=1,\ldots,\dNon$: $\quad\bmu_{\qDens(u_j)}\thickarrow 
\mu_{\qDens(\gammaSUBuj)}\,\bmu_{\qDens(\butilde_{j})}$
\item[] For $j=1,\ldots,\dLin+\dNon$:
\begin{itemize}
\item[] $\omegaOOO\thickarrow \be_j^T\XTyadj-\displaystyle{\sum_{j'=1}^{\dNon}}
\Big(\ZTXjd\be_j\Big)^T\bmu_{\qDens(u_{j'})}$
\item[] $\omegaOOO\thickarrow \mu_{\qDens(\betatilde_j)}\omegaOOO-\big(\XTX\be_j\big)_{-j}^T
\Big[\big(\bmu_{\qDens(\gammaSUBbeta)}\big)_{-j}\odot
\Big\{\Big(\bSigma_{\qDens(\bbetatilde)}\be_j\Big)_{-j}+\mu_{\qDens(\betatilde_j)} 
\Big(\bmu_{\qDens(\bbetatilde)}\Big)_{-j}\Big\}\Big]
$
\item[] $\mu_{\qDens(\gammaSUBbetaj)}\thickarrow
\expit\left(\logit(\rhoSUBbeta)-\smhalf\mu_{\qDens(1/\sigeps^2)}
\Big\{\big(\mu^2_{\qDens(\betatilde_j)}+
\sigma^2_{\qDens(\betatilde_j)}\big)\be_j^T\XTX\be_j-2\omegaOOO\Big\}\right)$
\end{itemize}
\item[] \textit{continued on a subsequent page $\ldots$}
\end{itemize}
\end{itemize}
\end{small}
\end{minipage}
\end{center}
\caption{\textit{Iterative determination of the optimal parameters according to
a mean field variational Bayes approximation of the posterior distributions  
for model (\ref{eq:mainModel}).}}
\label{alg:MFVB}
\end{algorithm}
%

\setcounter{algorithm}{2}
\begin{algorithm}[!th]
\begin{center}
\begin{minipage}[t]{160mm}
\begin{small}
\begin{itemize}
\item[]
\begin{itemize}
\item[] For $j=1,\ldots,\dNon$: $\quad\bmu_{\qDens(u_j)}\thickarrow 
\mu_{\qDens(\gammaSUBuj)}\,\bmu_{\qDens(\butilde_{j})}$
\item[] For $j=1,\ldots,\dNon$:
\begin{itemize}
\item[] $\omegaPPP\thickarrow\ZTyjadj-\ZTXj\Big(\bmu_{\qDens(\gammaSUBbeta)}\odot\bmu_{\qDens(\bbetatilde)}\Big)
-{\displaystyle\sum_{j'\ne j}^{\dNon}}\ZTZjjd\bmu_{\qDens(\bu_{j'})}$ 
\item[] $\bsigma^2_{\qDens(\butilde_j)}\thickarrow\bone_{K_j}\Big/\Big\{\mu_{\qDens(1/\sigeps^2)}
\mu_{\qDens(\gammaSUBuj)}\bwZj
+\mu_{\qDens(1/\sigma_{uj}^2)}\mu_{\qDens(b_{uj})}\bone_{K_j}\Big\}$
\item[] $\bmu_{\qDens(\butilde_j)}\thickarrow\mu_{\qDens(1/\sigeps^2)}
\Big(\mu_{\qDens(\gammaSUBuj)}\omegaPPP\Big)\odot\bsigma^2_{\qDens(\butilde_j)}$
\end{itemize}
\item[] For $j=1,\ldots,\dNon$:
\begin{itemize}
\item[] $\omegaQQQ\thickarrow \Vert\bmu_{\qDens(\butilde_j)}\Vert^2
+\bone_{K_j}^T\bsigma^2_{\qDens(\butilde_j)}$
\ \ \ ;\ \ \ $\mu_{\qDens(b_{uj})}\thickarrow\Big(\mu_{\qDens(1/\sigma_{uj}^2)}\omegaQQQ\Big)^{-1/2}$
\item[] $\lambda_{\qDens(\sigma_{uj}^2)}\thickarrow\mu_{\qDens(1/a_{uj})}+\smhalf\mu_{\qDens(b_{uj})}
\omegaQQQ$\ \ \ ;\ \ \ 
$\mu_{\qDens(1/\sigma_{uj}^2)}\thickarrow\kappa_{\qDens(\sigma_{uj}^2)}/\lambda_{\qDens(\sigma_{uj}^2)}$
\item[] $\lambda_{\qDens(a_{uj})}\thickarrow\mu_{\qDens(1/\sigma^2_{uj})}+(1/s_u^2)$
\ \ \ ;\ \ \ $\mu_{\qDens(1/a_{uj})}\thickarrow \kappa_{\qDens(a_{uj})}\big/\lambda_{\qDens(a_{uj})}$
\end{itemize}
\item[] For $j=1,\ldots,\dNon$: $\quad\bmu_{\qDens(u_j)}\thickarrow 
\mu_{\qDens(\gammaSUBuj)}\,\bmu_{\qDens(\butilde_{j})}$
\item[] For $j=1,\ldots,\dNon$:\\[-3ex]
\begin{itemize}
\item[] $\omegaRRR\thickarrow \ZTyjadj
-\ZTXj\big(\bmu_{\qDens(\gammaSUBbeta)}\odot\bmu_{\qDens(\bbetatilde)}\big)
-\displaystyle{\sum_{j'\ne j}^{\dNon}}\ZTZjjd\bmu_{\qDens(\bu_{j'})}$\\[0ex]
\item[] $\omegaSSS\thickarrow 
\bwZj^T\Big(\bmu_{\qDens(\butilde_j)}\odot\bmu_{\qDens(\butilde_j)}+\bsigma^2_{\qDens(\butilde_j)}\Big)
-2\bmu_{\qDens(\butilde_j)}^T\omegaRRR$
\item[] $\mu_{\qDens(\gamma_{uj})}\thickarrow \expit\Big(\logit(\rhoSUBu)
-\smhalf\mu_{\qDens(1/\sigeps^2)}\omegaSSS\Big)$
\end{itemize}
\item[] $\omegaTTT\thickarrow \bone_n\mu_{\qDens(\beta_0)}+
\bX\Big(\bmu_{\qDens(\gammaSUBbeta)}\odot\bmu_{\qDens(\bbetatilde)}\Big)
+\displaystyle{\sum_{j=1}^{\dNon}}\bZ_j\Big(\mu_{\qDens(\gammaSUBuj)}\bmu_{\qDens(\butilde_j)}\Big)$\\[-2ex]
\item[] If $\textbf{responseType}\ \mbox{is Gaussian}$ then\\[-2.5ex]
\begin{itemize}
\item[] $\bOmega_{\qDens(\gammaSUBbeta)}\thickarrow
\diag\big\{\bmu_{\qDens(\gammaSUBbeta)}\odot(\bone-\bmu_{\qDens(\gammaSUBbeta)})\big\}
+\bmu_{\qDens(\gammaSUBbeta)}\bmu_{\qDens(\gammaSUBbeta)}^T$
\\[-2ex]
\item[]
$\lambda_{\qDens(\sigeps^2)}\thickarrow\mu_{\qDens(1/\aSUBeps)}+\smhalf\Vert\by-\omegaTTT\Vert^2
+\smhalf\,n\sigma^2_{\qDens(\beta_0)}$\\[1ex]
\null
$\qquad\qquad\qquad 
+\smhalf\tr\Big[\bX^T\bX\Big\{\bOmega_{\qDens(\gammaSUBbeta)}\odot\Big(\bSigma_{\qDens(\bbetatilde)}
+\bmu_{\qDens(\bbetatilde)}\bmu_{\qDens(\bbetatilde)}^T\Big)\Big\}\Big]$\\[1ex]
\null$\qquad\qquad\qquad -\smhalf\tr\Big\{\bX^T\bX
\Big(\bmu_{\qDens(\gammaSUBbeta)}\odot\bmu_{\qDens(\bbetatilde)}\Big)
\Big(\bmu_{\qDens(\gammaSUBbeta)}\odot\bmu_{\qDens(\bbetatilde)}\Big)^T\Big\}$\\[1ex]
\null$\qquad\qquad\qquad +\smhalf{\displaystyle\sum_{j=1}^{\dNon}}\bwZj^T
\Bigg(\mu_{\qDens(\gammaSUBuj)}
\Big[
\bsigma^2_{\qDens(\butilde_j)}
+\big\{1-\mu_{\qDens(\gammaSUBuj)}\big\}
\bmu_{\qDens(\butilde_j)}\odot\bmu_{\qDens(\butilde_j)}\Big]\Bigg)$
\\
[0ex]
\item[] $\mu_{\qDens(1/\sigeps^2)}\thickarrow\kappa_{\qDens(\sigeps^2)}\Big/\lambda_{\qDens(\sigeps^2)}$
\ ;\ $\lambda_{\qDens(\aSUBeps)}\thickarrow\mu_{\qDens(1/\sigeps^2)}
+(1/\sSUBeps^2)$\ ;\ $\mu_{\qDens(1/\aSUBeps)}\thickarrow \kappa_{\qDens(\aSUBeps)}\Big/\lambda_{\qDens(\aSUBeps)}$
\end{itemize}
\item[] If $\textbf{responseType}\ \mbox{is Bernoulli}$ then\\[-3ex]
\begin{itemize}
\item[] $\mu_{\qDens(1/\sigeps^2)}\thickarrow 1$\ \ ;\ \  
$\bmu_{\qDens(\bc)}\thickarrow\omegaTTT+(2\by-\bone_n)\odot\zeta'\big((2\by-\bone_n)\odot \omegaTTT\big)$
\item[] $\yToneadj\thickarrow \bmu_{\qDens(\bc)}^T\bone_n$\ \ ;\ \  $\XTyadj\thickarrow \bX^T\bmu_{\qDens(\bc)}$
\ \ ;\ \  $\ZTyadj\thickarrow \bZ^T\bmu_{\qDens(\bc)}$
\end{itemize}
\end{itemize}
\item[] until the relative change in the $\log\punder(\by;\qDens)$ 
is below $\varepsilon_{\tiny\mbox{toler}}$.
\item[] Outputs: All $\qDens$-density parameters.
\end{itemize}
\end{small}
\end{minipage}
\end{center}
\caption{\textbf{continued.}\ \textit{This is a continuation of the description of this algorithm that
commences on a preceding page.}}
\end{algorithm}
%

\subsection{Model Selection Strategies}\label{sec:modSelecStrat}

Essential components of our Bayesian generalized additive model selection methodology 
are rules, based on the posterior distributions of relevant parameters, for 
deciding whether an effect is zero, linear or non-linear. In practice, either
the Markov chain Monte Carlo samples or mean field variational Bayes $\qDens$-densities
are used for approximate posterior-based decision making. However, we will
describe our strategies in terms of exact posterior distributions -- starting
with the zero versus linear effect decision.

\subsubsection{Deciding Between an Effect Being Zero or Linear}\label{sec:ZeroVsLin}

Let $\beta$ be a generic regression coefficient attached to one of the 
$\bxLinj$ predictors. According to our models, $\beta=\gammaSUBbeta\,\betatilde$
where $\gammaSUBbeta$ is binary and $\betatilde$ is continuous. Therefore
$$P(\beta=0|\by)=P(\gammaSUBbeta=0|\by)=1-E(\gammaSUBbeta|\by),$$
and the posterior mean of $\gammaSUBbeta$ can be used to decide between
hypotheses $H_0:\beta=0$ and $H_1:\beta\ne0$. A natural rule is to accept $H_0$ 
if and only if 
$$P(\beta=0|\by)>\smhalf\quad\mbox{which is equivalent to}\quad E(\gammaSUBbeta|\by)\le\smhalf.$$
However, in the interests
of parsimony, less stringent rules are worth considering. Rather than
exclusively thresholding $E(\gammaSUBbeta|\by)$ at $\smhalf$, we also 
consider a family of rules indexed by a threshold parameter $\tau\in(0,1)$.
After fixing $\tau$  our strategy for deciding between an effect being zero or linear is
$$\mbox{the effect is zero if}\ E(\gammaSUBbeta|\by)\le1-\tau,\ 
\mbox{otherwise the effect is linear}.$$
According to this definition of the threshold parameter, lower values of $\tau$ lead to sparser
fits. 

\subsubsection{Deciding Between an Effect Being Zero, Linear or Non-Linear}\label{sec:ZeroVsLinVsNonlin}

Now let $\beta$ be a generic linear coefficient and $\bu$ be a generic $K\times1$ spline 
coefficient vector attached to one of the $\bxNonj$ predictors.
Since $\bu=\gammaSUBu\butilde$, where the entries of $\gammaSUBu$
are binary and the entries of $\butilde$ are continuous, 
$$P(\bu=\bzero|\by)=P(\gammaSUBu=0|\by)=1-E(\gammaSUBu|\by).$$
Therefore, after fixing $\tau$, our strategy for deciding between an effect being zero, 
linear or non-linear is:
\begin{eqnarray*}
&&\mbox{the effect is zero if}\ 
\mbox{max}\big\{E(\gammaSUBbeta|\by),E(\gammaSUBu|\by)\big\}\le 1-\tau,\\[1ex]
&&\mbox{the effect is linear if}\ E(\gammaSUBbeta|\by)>1-\tau\ \mbox{and}\
E(\gammaSUBu|\by)\le1-\tau,\\[1ex]
&&\mbox{otherwise the effect is non-linear.}
\end{eqnarray*}

It is apparent from these rules that the parameter $\tau\in(0,1)$ controls
the degree of sparsity in the selected model, with lower values of $\tau$ producing
sparse fits. Hence, we refer to $\tau$ as the \emph{sparsity threshold parameter}. 
In practice, various values of $\tau$ can be contemplated but for a completely 
automatic model selection a good default choice is desirable. 
We confront this problem in the next subsubsection.

\subsubsection{Choice of Default Values for the Sparsity Threshold Parameter}\label{sec:choiceTau}

Among the family of rules indexed by the sparsity threshold parameter 
$\tau\in(0,1)$, an important practical question is that of recommending
a default value for $\tau$.
To aid such a recommendation, we simulated data sets from both Gaussian and Bernoulli response 
generalized additive models with $\dNon=30$ continuous predictors. 
Ten of the predictors had a zero effect, $10$ had linear effects
with random generated coefficients, and  $10$ had non-linear effects.
Each of the predictors were generated from independent Normal distributions. 
The non-linear effects corresponded to quintic polynomials with 
randomly generated coefficients. 
Each replication involved the generation of new coefficients.
The sample sizes varied over $n\in\{500,1000,2000\}$ and, for the Gaussian response case, 
the error standard deviations varied over $\sigma_{\varepsilon}\in\{0.25,0.5,1,2\}.$
For each combination of sample size and error standard deviation $100$ 
data sets were generated. Fitting was carried out using both 
Algorithm \ref{alg:MCMC} with $\Nwarm=\Nkept=1000$ and Algorithm \ref{alg:MFVB}
with $\varepsilon_{\tiny\mbox{toler.}}=10^{-8}$. Model selection was applied 
according to the rules of Section \ref{sec:modSelecStrat}
with $\tau\in\{0.1,0.3,0.5,0.7,0.9\}$. The performance measure was 
misclassification rate for the $30$ candidate predictors being classified into 
one of three classes: zero effect, linear effect and non-linear effect.

\begin{figure}[!t]
\centering
\includegraphics[width=1.0\textwidth]{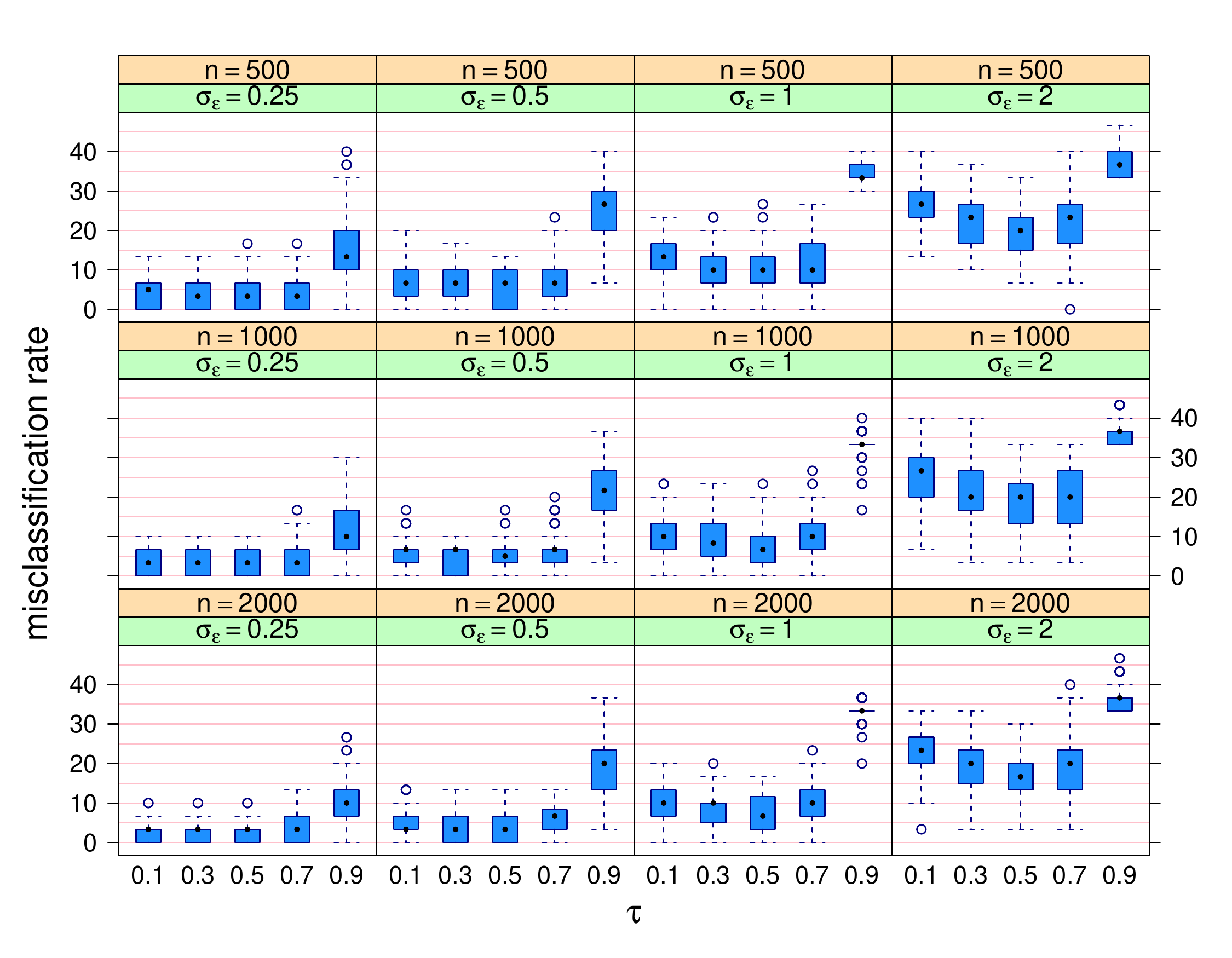}
\caption{
\textit{Side-by-side boxplots of the misclassification rates
for the Markov chain Monte Carlo Algorithm \ref{alg:MCMC}
for the simulation study described in the text. Each panel corresponds
to a different combination of sample size and error standard deviation.
Within each panel, the side-by-side boxplots compare misclassification
rate as a function of the threshold parameter $\tau$.}}
\label{fig:choiceTauMCMCgauss}
\end{figure}

Figure \ref{fig:choiceTauMCMCgauss} displays the misclassification rate
data for Algorithms \ref{alg:MCMC} from $100$ simulation replications.
Each panel corresponds to a different combination of sample size and error 
standard deviation. Within each panel, side-by-side boxplots of the 
misclassification rate are shown as a function of $\tau$. For low
noise levels there is not much of a difference, but for $\sigeps\ge1$
it is advantageous to have $\tau$ equal to the natural choice of $0.5$.
Note, however, that this recommendation is necessarily limited due to 
being based on a single simulation study.

\begin{figure}[!t]
\centering
\includegraphics[width=1.0\textwidth]{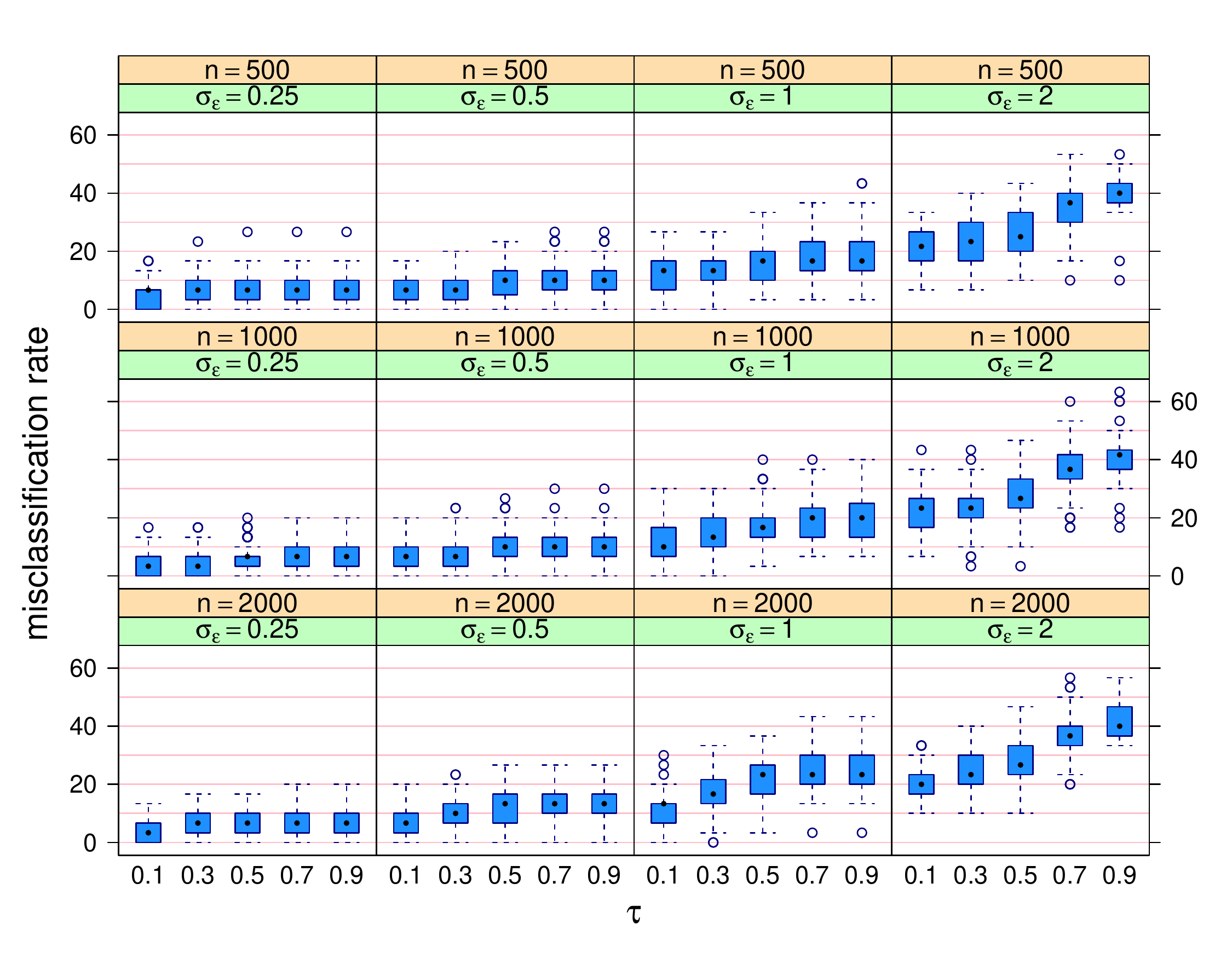}
\caption{
\textit{Side-by-side boxplots of the misclassification rates
for the mean field variational Bayes Algorithm \ref{alg:MFVB} for the 
simulation study described in the text. Each panel corresponds
to a different combination of sample size and error standard deviation.
Within each panel, the side-by-side boxplots compare misclassification
rate as a function of the threshold parameter $\tau$.}}
\label{fig:choiceTauMFVBgauss}
\end{figure}

The analogous results for the mean field variational Bayes approach
of Algorithm \ref{alg:MFVB} are shown in Figure \ref{fig:choiceTauMFVBgauss}.
This time the boxplots indicate better performance for $\tau<0.5$.
We conjecture that mean field approximations have a detrimental effect 
on the $\tau=\smhalf$ decision rules and, for reasons yet to be
understood, are somewhat remedied by setting $\tau$ to be a lower
value such as $0.1$. Additional checks, not shown here, indicate the
the classification performance gets worse for $\tau$ smaller than $0.1$
for this simulation set-up. Acknowledging the limitations of a single simulation 
study, our recommended default for $\tau$ in the mean field variational Bayes
case is $\tau=0.1$.

We also ran simulation studies for the Bernoulli response case, 
with a similar design to the Gaussian study. The recommendations
of $\tau=0.5$ for Markov chain Monte Carlo and $\tau=0.1$ for 
mean field variational Bayes were also supported by that study.

Additional simulation studies, involving an alternative evaluation
metric and hyperparameter sensitivity checks, are given in 
Section \ref{sec:addSimRes}. These studies do not alter any
of our recommendations concerning the choice of $\tau$.

\subsection{Package in the \textsf{R} Language}\label{sec:Rpackage}

The \textsf{R} package \textsf{gamselBayes} (He \myand Wand, 2023)
implements Algorithms \ref{alg:MCMC} and \ref{alg:MFVB} and provides
tabular and graphical summaries of selected generalized additive models.
Speed is enhanced via \texttt{C++} implementation of the loops in the
two algorithms. The \textsf{gamselBayes} package is available on the Comprehensive \textsf{R} 
Archive Network (\texttt{https://www.R-project.org}). The \textsf{gamselBayes} package 
is accompanied  by a vignette which provides fuller details on its use. 
The vignette PDF file is opened via the command \texttt{gamselBayesVignette()}. 

\section{Comparative Performance}\label{sec:performance}

We ran a second simulation study to assess comparative performance of the new methodology
with respect to some of the other existing approaches to three-category generalized additive 
model selection. The simulation design was the same as that described in Section \ref{sec:choiceTau}. 
In keeping with the findings of that section, in Algorithm \ref{alg:MCMC} the threshold parameter 
was set to $\tau=0.5$ and for Algorithm \ref{alg:MFVB} it was set to $\tau=0.1$. 

The other approaches considered were those used by the \textsf{R} packages:
\begin{enumerate}
\item \textsf{spikeSlabGAM} (Scheipl, 2022), which is a Bayesian approach that is described
in Scheipl \textit{et al.} (2012). Details on use of the \textsf{spikeSlabGAM} package are 
given in Scheipl (2011).
\item \textsf{gamsel} (Chouldechova \myand Hastie, 2022), which implements the frequentist approach 
described in Chouldechova \myand Hastie (2015). The package's main function, \texttt{cv.gamsel()},
computes a family of generalized additive model fits over a grid of regularization parameter values.
For selection of a single model, \texttt{cv.gamsel()} provides the option of minimizing a $k$-fold
cross-validation function over the grid.
\end{enumerate}

The Bernoulli response versions of these approaches involve the logit
link function, rather than the probit link function used by
Algorithms \ref{alg:MCMC} and \ref{alg:MFVB}.
This necessitated use of the appropriate inverse link
transformation for the generation of binary response data in
this simulation study.

In the case of \textsf{spikeSlabGAM}, we used the default call to its \texttt{spikeSlabGAM()} function.
The model having highest posterior probability in the \texttt{spikeSlabGAM()} output object was selected.
The essential difference between \textsf{spikeSlabGAM} and Algorithms \ref{alg:MCMC} and \ref{alg:MFVB}
is the form of the prior distributions imposed on the coefficients for the 
linear and spline components. In the notation of Section \ref{sec:modLinCoeff}, 
\textsf{spikeSlabGAM} replaces (\ref{eq:TwinCities}) by
\begin{equation}
\pDens(\beta|\sigmaSUBbeta,\rhoSUBbeta)=
\frac{\rhoSUBbeta\exp\big\{-\beta^2/(2\sigmaSUBbeta^2)\big\}}{(2\pi\sigmaSUBbeta^2)^{1/2}}
+\frac{(1-\rhoSUBbeta)\exp\big[-\beta^2/\{2(v_0\sigmaSUBbeta)^2\}\big]}
{\{2\pi(v_0\sigmaSUBbeta)^2\}^{1/2}}
\quad\mbox{where}\quad v_0\ll 1
\label{eq:MoscowIdaho}
\end{equation}
with $v_0$ having a default value of $0.00025$. Note (\ref{eq:MoscowIdaho}) 
is an alternative to the ``spike-and-slab'' prior used by (\ref{eq:TwinCities}),
with the ``slab'' being Gaussian rather than Laplacian and the default 
``spike'' being a $N(0,0.00025^2)$ mass rather than the point mass at zero.
For spline coefficient vectors, the alternative to (\ref{eq:DontArgue}) 
used by \textsf{spikeSlabGAM} is an extension of (\ref{eq:MoscowIdaho}) 
that is described by Figure 1 of Scheipl (2011) and accompanying text.
For Gaussian response models \textsf{spikeSlabGAM} uses Gibbs sampling,
but requires Metropolis-Hastings sampling for non-Gaussian responses.

\begin{figure}[!t]
\centering
\includegraphics[width=1.0\textwidth]{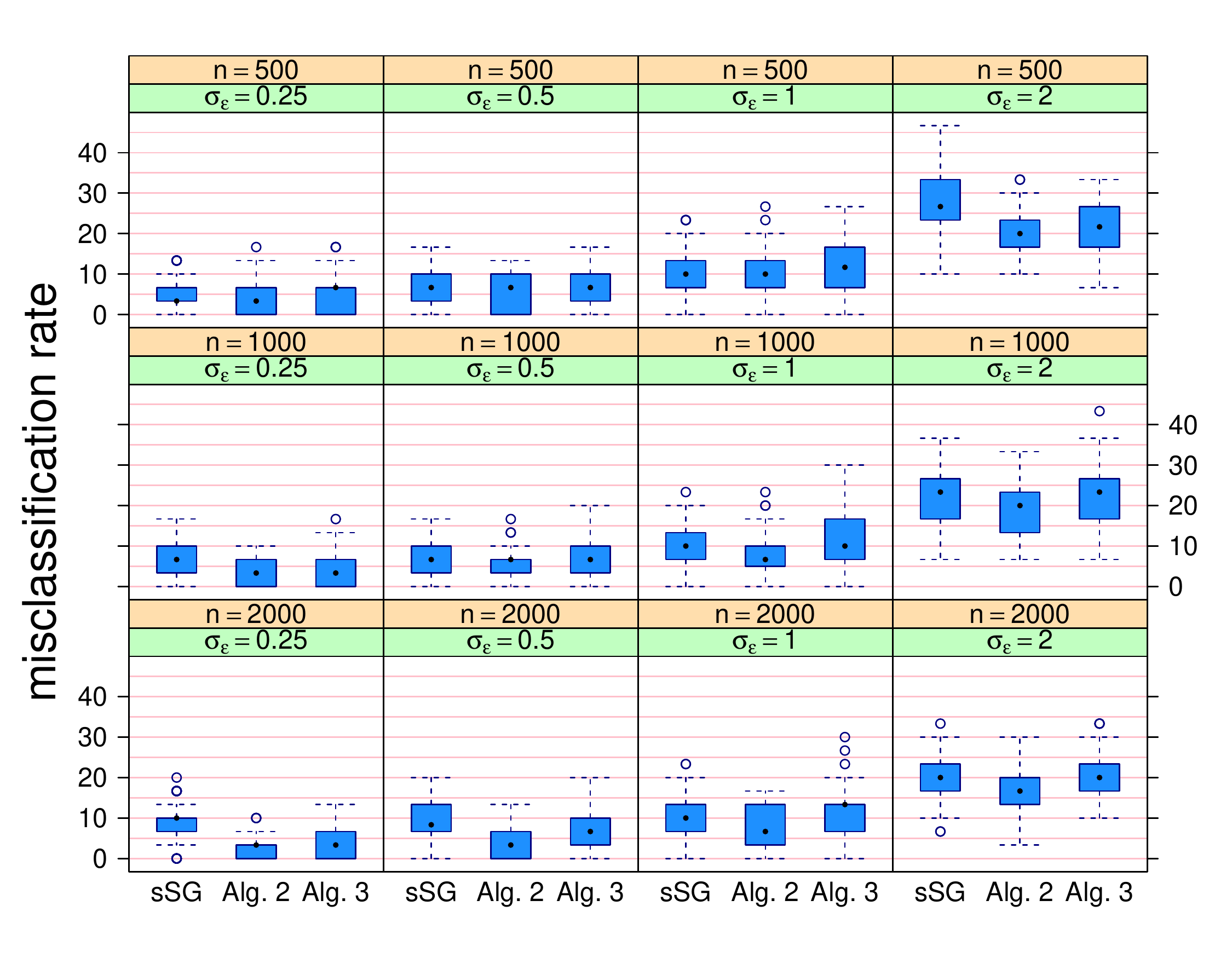}
\caption{
\textit{Side-by-side boxplots of the misclassification rates
for the comparative performance simulation study described in the text in the
case of the response variable being Gaussian. Each panel corresponds to a different 
combination of sample size and error standard deviation. Within each panel, the 
side-by-side boxplots compare misclassification rate across each of three methods:} 
\textsf{spikeSlabGAM} \textit{with default settings (sSG), Algorithm \ref{alg:MCMC} (Alg. 2)
and Algorithm \ref{alg:MFVB} (Alg. 3)}.}
\label{fig:simulBoxpGauss}
\end{figure}

\begin{figure}[!t]
\centering
\includegraphics[width=0.85\textwidth]{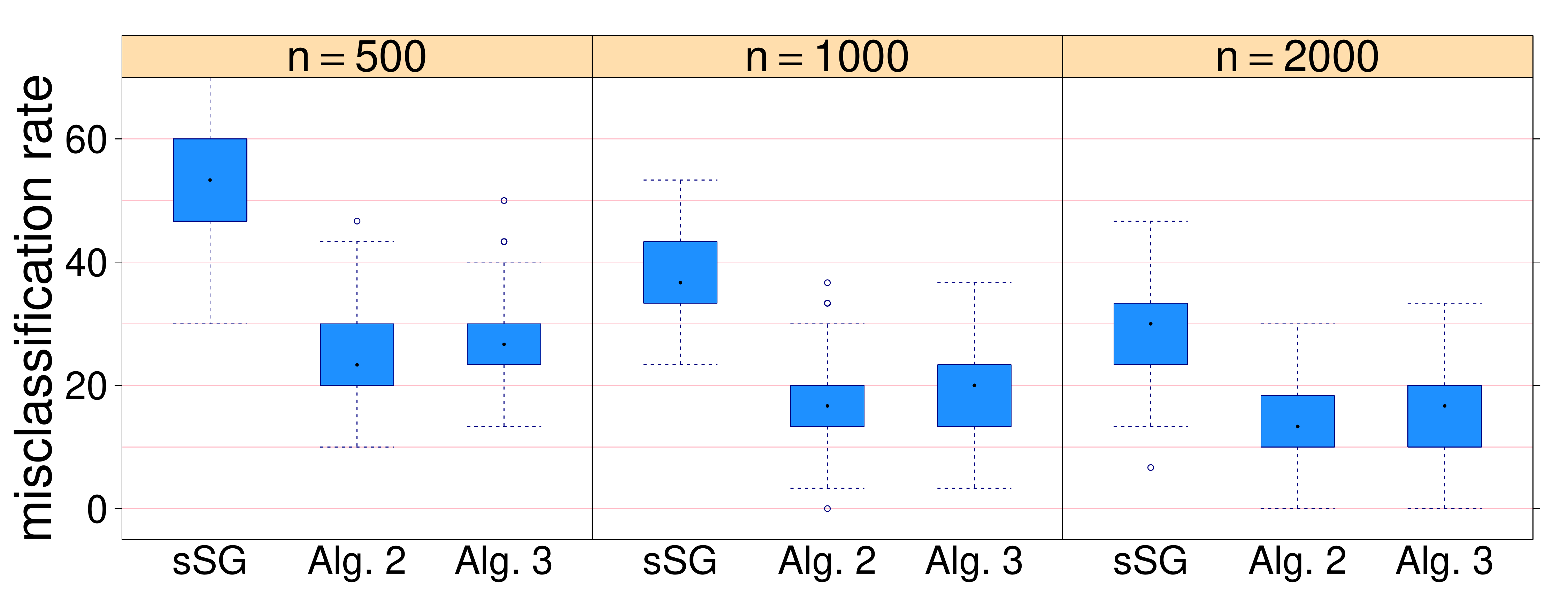}
\caption{
\textit{Side-by-side boxplots of the misclassification rates
for the comparative performance simulation study described in the text in the
case of the response variable being binary. Each panel corresponds to a different 
combination of sample size and error standard deviation. Within each panel, the 
side-by-side boxplots compare misclassification rate across each of three methods:} 
\textsf{spikeSlabGAM} \textit{with default settings (sSG),
Algorithm \ref{alg:MCMC} (Alg. 2) and Algorithm \ref{alg:MFVB} (Alg. 3)}.}
\label{fig:simulBoxpBern}
\end{figure}

Preliminary checks revealed that default regularization grid used by \texttt{cv.gamsel()} did not lead
to very good three-category classification performance, with the cross-validation mean function often being 
monotonic rather than U-shaped. To circumvent this apparent default grid problem, with respect
to the three-category misclassification rate, we experimented with its choice and found that 
geometric sequence of size $50$ between $0.01$ and $2$ usually lead to U-shaped cross-validation 
mean functions for the simulation settings. This regularization grid was used throughout the 
comparative performance simulation study  with $10$-fold cross-valid\-ation for model selection. 
Two cross-validation-based choices
were considered: the regularization parameter matching the absolute minimum of the mean values, 
and largest regularization parameter value such that mean minus one standard deviation is below the
absolute minimum. However, after running the simulation study it was found that the three-category 
misclassification rates for the \textsf{gamsel} approaches were considerably higher than the other 
approaches since it has a tendency to choose larger models. Given this poor performance
for misclassification rate, relative to the other methods in the study, the \textsf{gamsel} results
are excluded from the upcoming graphical summaries (Figures \ref{fig:simulBoxpGauss} 
and \ref{fig:simulBoxpBern}).

Figure \ref{fig:simulBoxpGauss} shows the misclassification rates for Algorithms \ref{alg:MCMC} 
and \ref{alg:MFVB} in comparison with the default version of the \textsf{spikeSlabGAM} approach
as side-by-side boxplots for the Gaussian response case. In the lower error standard deviation 
situations, all have similar performance. The fast variational approach of Algorithms \ref{alg:MFVB} 
is seen to have lower accuracy when the noise level is higher. This degradation in performance
needs to be mitigated against run time, which is addressed later in this section.

The binary response simulation results are shown in Figure \ref{fig:simulBoxpBern}.
Algorithms \ref{alg:MCMC} and \ref{alg:MFVB} are seen to have better three-category
classification performance compared with \textsf{spikeSlabGAM} for the binary 
response simulation study.

Lastly, we report on the computing times for the four approaches.
Specifically, these are elapsed times in seconds for each generalized additive model 
selection on a \textsf{MacBook Air} laptop computer with 16 gigabytes of memory and 
a 3.2 gigahertz processor. Algorithms \ref{alg:MCMC} and \ref{alg:MFVB} 
were implemented using the \textsf{Rcpp} interface (Eddelbuettel \myand Fran{\c{c}}ois, 2011)
to the \texttt{C++} language. The Markov chain Monte Carlo sample size values corresponded
to the \textsf{spikeSlabGAM} and \textsf{gamselBayes} defaults of $1,500$ and $2,000$ 
respectively. Table \ref{tab:computTimes} lists the 10th, 50th and 90th percentile number of seconds
for each approach across all settings and replications.

\begin{table}[!t]
\begin{center}
\begin{tabular}{rrrrr}
\hline\\[-1.8ex]
&\textsf{gamsel} & \textsf{spikeSlabGAM} &  Algorithm \ref{alg:MCMC} & Algorithm \ref{alg:MFVB} \\[0.1ex]
\hline\\[-1.5ex]
10th percentile &  3.69     &     84.9     &   1.78     &   0.326\\
50th percentile &  8.12     &    167.0     &   2.14     &   0.466\\
90th percentile & 17.50     &    339.0     &   3.03     &   0.768\\
\hline
\end{tabular}
\end{center}
\caption{\textit{10th, 50th and 90th percentiles for the number of seconds required for each 
generalized additive model selection approach across all settings and replications for the comparative 
performance simulation study.}}
\label{tab:computTimes}
\end{table}

It is apparent from Table \ref{tab:computTimes} that, despite exhibiting very good classification,
\textsf{spikeSlabGAM} is comparatively slow and does not scale well to large
problems. Algorithm \ref{alg:MCMC} took less than around 3 seconds for 90\% of the fits
in the simulation study. The faster variational approach of Algorithm \ref{alg:MFVB} 
only required less than a second of computing time for most of the fits. Therefore, the new 
approaches have very good scalability for the generalized additive model selection problem.

The impact of sample size and number of candidate predictors on 
computing times for Algorithms \ref{alg:MCMC} and \ref{alg:MFVB}
is investigated in Section \ref{sec:detCompTimeRes}.

\section{Data Illustrations}\label{sec:dataIllus}

We finish off with two illustrations for actual data. Both illustrations
involve binary responses. The first one is a relative small problem, where Markov 
chain Monte Carlo fitting of the binary response adjustment of (\ref{eq:mainModel})
is quick. The second example involves a much bigger data set, and mean field variational Bayes 
offers relatively fast model selection.

\subsection{Application to Mortgage Applications Data}

Data originating from the Federal Bank of Boston, U.S.A., has
$2,380$ records on mortgage applications, and is available
in the \textsf{R} data package \textsf{Ecdat} (Croissant, 2022) 
as a data frame titled \texttt{Hmda}. The response variable is the indicator
of whether the mortgage application was denied.
After conversion of each of the categorical variables to 
indicator form there are 20 candidate predictors. Fourteen of these candidate
predictors are binary, so can only be considered as having a zero or linear
effect. The remaining four predictors are continuous, and three of these
were considered as having zero, linear or non-linear effects. One
of them, corresponding to the unemployment rate of the industry corresponding to 
the applicant's occupation, has only 10 unique values and penalized spline
models have borderline viability. Therefore, the effect of this predictor was
restricted to zero versus linear.

Application of Algorithm \ref{alg:MCMC} and the effect type estimation rules 
of Section \ref{sec:modSelecStrat} with $\tau=0.5$ led to the estimated effect types 
listed in Table \ref{tab:BostMortEffTypes}. Markov chain Monte Carlo sampling 
involved a warm-up of length $1,000$ and $1,000$ retained samples used for inference. 
Chain diagnostic graphics, including trace, lag-1 and autocorrelation function plots, 
indicated good convergence. The vignette attached to the \textsf{gamselBayes} package
includes these diagnostic graphics.

\begin{table}[!t]
\begin{center}
\begin{tabular}{lll ll}
\hline
candidate predictor        &est. type & &candidate predictor     &est. type\\
\hline
bad public credit record?  & linear   & & credit score of 3?         & zero \\
denied mortgage insurance? & linear   & & credit score of 4?    & zero    \\
applicant self-employed?   & linear     & & credit score of 5?    & zero      \\
applicant single?          & linear   & & mortgage credit score of 1? 	& zero      \\ 
applicant black?           & linear   & & mortgage credit score of 2?    & zero      \\ 
property a condominium?    & zero     & & mortgage credit score of 3?	& zero       \\ 
unemploy. rate applic. indus.  & zero & & debt payments/income ratio	& non-linear  \\
credit score of 1?        & linear    & & housing expenses/income ratio& zero         \\
credit score of 2?        & linear    & & loan size/property value ratio& non-linear \\
 \hline
\end{tabular}
\end{center}
\caption{\textit{Each of the candidate predictors for the Boston mortgage example and the
estimated effect type from application of Algorithm \ref{alg:MCMC} and effect type estimation 
rules of Section \ref{sec:modSelecStrat}. The candidate predictors with question marks correspond
to binary indicator variables. The abbreviation ``unemploy. rate applic. indus.'' stands for
the unemployment rate of the industry corresponding to the applicant's occupation.}}
\label{tab:BostMortEffTypes}
\end{table}

As is apparent from Table \ref{tab:BostMortEffTypes}, the selected model has 7 linear
effects, 2 non-linear effects and 9 candidate predictors discarded. Table 
\ref{tab:BostMortLinEffecs} provides estimation and inferential summaries for the
linear effects.

\begin{table}[!t]
\begin{center}
\begin{tabular}{lrr}
 predictor &       posterior mean  &   95\% credible interval\\[1ex]
\hline
indicator of bad public credit record     &     $0.7350$   &     $(0.4926,0.9848)$    \\[0.5ex]
indicator of denied mortgage insurance    &     $2.7620$   &     $(2.1426,3.5172)$    \\[0.5ex]
indicator of applicant being single       &     $0.1370$   &     $(0.0000,0.3417)$    \\[0.5ex] 
indicator of applicant being black        &     $0.3461$   &     $(0.0842,0.5404)$    \\[0.5ex]  
indicator of applicant being self-employed&     $0.1703$   &     $(0.0000,0.4363)$    \\[0.5ex]  
indicator of credit score equalling 1     &    $-0.6906$   &     $(-0.8980,-0.4513)$ \\[0.1ex] 
indicator of credit score equalling 2     &    $-0.3238$   &     $(-0.5869,0.0000)$ \\[0.1ex] 
\hline
\end{tabular}
\end{center}
\caption{\textit{Approximate posterior means and approximate 95\% credible intervals for the coefficients
of each of the selected linear fits based on the Markov chain Monte Carlo samples generated
from Algorithm \ref{alg:MCMC} for the Boston mortgages example.}}
\label{tab:BostMortLinEffecs}
\end{table}

Table \ref{tab:BostMortEffTypes} shows an applicant having bad public credit
record is more likely to have their mortgage application denied, which is
in keeping with financial commonsense. Of potential interest from a social justice
standpoint is the significant effects on denial probability for applicants that
are either black or single.

\begin{figure}[!t]
\centering
\includegraphics[width=1.0\textwidth]{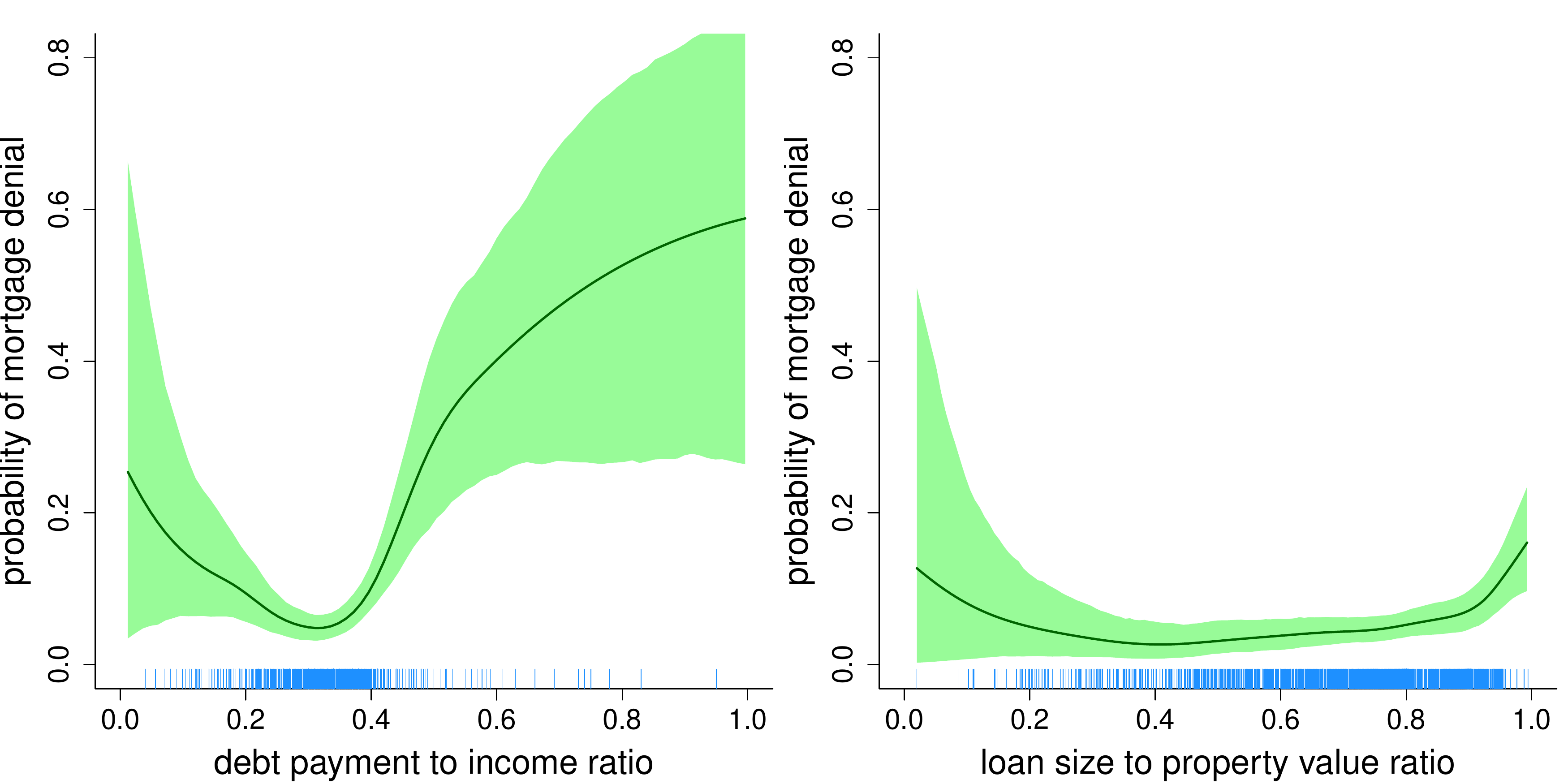}
\caption{
\textit{The two estimated non-linear effects for the Boston mortgage example
from application of Algorithm \ref{alg:MCMC} and effect type estimation rules of Section 
\ref{sec:modSelecStrat} with $\tau=0.5$. Each curve is the slice of estimated probability 
of mortgage denial as a function of the predictor, with all other selected predictors 
set to their median values. The shaded region corresponds to pointwise approximate
95\% credible intervals.}}
\label{fig:BostonMortNonlin}
\end{figure}

Figure \ref{fig:BostonMortNonlin} shows the two effects have non-linear effects in
the selected model. The effect of debt payment to income ratio is quite a striking
non-monotonic curve.

\subsection{Application to Car Auction Data}\label{sec:carAuction}

During 2011-2012 the \textsf{kaggle} Internet platform (\texttt{https://www.kaggle.com})
hosted a classification competition involving training data consisting of $49$ variables
on $72,983$ cars purchased at automobile auctions by automobile dealerships in U.S.A.
The title of the competition was ``Don't Get Kicked!''. A version of the data in
which all categorical variables have been converted to binary variable indicator form
is stored in the data frame \texttt{carAuction} within the \textsf{R} package
\textsf{HRW} (Harezlak \textit{et al.}, 2021). The response variable is the indicator
of whether the car purchased at auction by the dealership had serious problems
that hinder or prevent it being sold. For short, we refer to such a car as a ``bad buy''.
Forty-four of the candidate predictors are binary. The other $5$ candidate predictors are 
continuous. However, the age at sale variable has only $10$ unique values. For the same reasons 
given for the unemployment rate variable considered in the Boston mortgages example, we 
exclude age at sale from having a non-linear effect.

Since this generalized additive model selection problem involves a relatively large sample
size and number of candidate predictors, we use it to illustrate the fast variational approach 
corresponding to Algorithm \ref{alg:MFVB}. The mean field variational Bayes iterations
described there were iterated until the relative change in the approximate marginal 
log-likelihood fell below $10^{-8}$. On the second author's \textsf{MacBook Air}
laptop, with a 3.2 gigahertz processor and 16 gigabytes of random access memory, 
mean field Bayes variational fitting took 11 seconds. 
The rules of Section \ref{sec:modSelecStrat}
were applied with $\tau=0.1$. This resulted in $19$ predictors
being selected as having a linear effect and $3$ predictors
having non-linear effects. Twenty-seven of the 
$49$, or $55\%$, of candidate predictors were discarded.

Table \ref{tab:carAuctLinEffs} provides estimation and inferential summaries for the
linear effects coefficients. Most of the predictor effects are intuitive, such as 
older cars being more likely to be a bad buy and presence of wheel covers lowering the 
bad buy probability. Some of them, such as the effect of cars
being purchased in particular states, are more intriguing.

\begin{table}[!t]
\begin{center}
\begin{tabular}{lrr}
 predictor                        & posterior mean      &   95\% credible interval\\[1ex]
\hline                          
indic. made in U.S.A.             & $-0.04605$         & $(-0.06925, -0.02323)$\\[0.1ex] 
age at sale (years)               & $0.09344$          & $(0.08771,0.09896)$   \\[0.1ex] 
indic. color is red               & $0.05130$          & $(0.02498,0.07713)$    \\[0.1ex] 
indic. make is Chevrolet          & $-0.1103$           & $(-0.1358,-0.08426)$   \\[0.1ex] 
indic. make is Chrysler           & $0.09517$          & $(0.06910,0.1217)$      \\[0.1ex] 
indic. make is Dodge              & $-0.03093$         & $(-0.05385,-0.006961)$ \\[0.1ex] 
indic. purchased online           & $-0.06229$         & $(-0.1117,0.0000)$          \\[0.1ex]
acquisition price (U.S. dollars)  & $-6.014\times10^{-6}$ & $(-9.352,-2.603)\times10^{-6}$\\[0.1ex]
indic. purchased in 2010          & $0.1085$           & $(0.09290,0.1242)$    \\[0.1ex] 
indic. purch. in Florida          & $-0.1173$          & $(-0.1391,-0.09529)$  \\[0.1ex]      
indic. purch. in North Carolina   & $-0.1074$          & $(-0.1326,-0.08190)$   \\[0.1ex] 
indic. purch. in Texas            & $0.09706$          & $(0.07799,0.1162)$     \\[0.1ex] 
indic. medium-sized vehicle       & $-0.07368$         & $(-0.09095,-0.05622)$\\[0.1ex] 
indic. sports utility vehicle     & $0.1899$           & $(0.1646,0.2153)$\\[0.1ex] 
indic. manual transmission        & $-0.1574$          & $(-0.1970,-0.1168)$\\[0.1ex] 
indic. trim level is `Bas'        & $0.05521$          & $(0.0355,0.07477)$\\[0.1ex] 
indic. trim level is `LS'         & $-0.06153$         & $(-0.09180,-0.03176)$\\[0.1ex] 
indic. has alloy wheels           & $-1.513$           & $(-1.546,-1.480)$\\[0.1ex] 
indic. has wheel covers           & $-1.585$           & $(-1.619,-1.551)$    \\[0.1ex] 
\hline
\end{tabular}
\end{center}
\caption{\textit{Approximate posterior means and approximate 95\% credible intervals 
for the coefficients of each of the selected linear fits based on the mean field 
variational Bayes optimal $\qDens$-densities obtained from Algorithm \ref{alg:MFVB} 
for the car auction example.}}
\label{tab:carAuctLinEffs}
\end{table}

\begin{figure}[!t]
\centering
\includegraphics[width=1.0\textwidth]{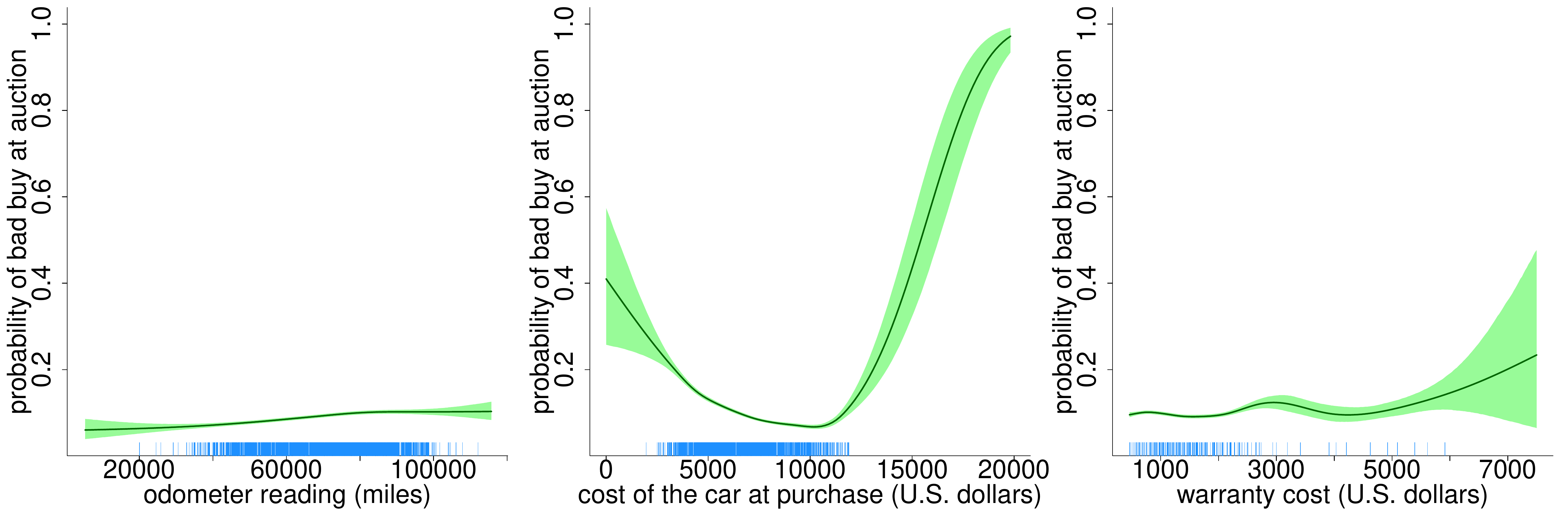}
\caption{
\textit{The estimated non-linear effects for the car auction example
from application of Algorithm \ref{alg:MFVB} and effect type estimation rules of Section 
\ref{sec:modSelecStrat} with $\tau=0.1$. The curves are slices of estimated probability 
of bad buy as a function of the predictor, with all other selected predictors 
set to their median values. The shaded regions correspond to pointwise 
approximate 95\% credible intervals, but are subject to considerable
mean field approximation error. The rug at the base of each plot is based on
a random sample of $2,500$ cars.}}
\label{fig:carAuctNonlin}
\end{figure}

Figure \ref{fig:carAuctNonlin} shows the three selected non-linear effects, 
which are the impacts of the probability of a bad buy as functions of the 
odometer reading in miles, acquisition cost paid for the car at the time 
of purchase in U.S. dollars and warranty cost in U.S. dollars.
The middle panel of Figure \ref{fig:carAuctNonlin} shows  that a cost of 
about $10,000$ U.S. dollars is best, and that the probability of bad buy 
increases when the cost deviates away from this amount.
The shaded regions of Figure \ref{fig:carAuctNonlin} corresponds to pointwise 
approximate 95\% credible intervals. However, for a binary response model such
as this, there is considerable mean field approximation error which tends to make
the credible intervals overly narrow.

As a type of check, we also applied the Markov chain Monte Carlo Algorithm \ref{alg:MCMC} to
the same data set. This resulted in 14 of the 19 predictors in Table \ref{tab:carAuctLinEffs}
being selected. Three predictors not selected by Algorithm \ref{alg:MFVB},
such as indicators of the auction provider, were selected by Algorithm \ref{alg:MCMC}.
The odometer reading predictor was estimated to have a non-linear effect by
Algorithm \ref{alg:MFVB}, but to have a linear effect by Algorithm \ref{alg:MCMC}.
In summary, Algorithm \ref{alg:MFVB} selected 22 predictors whilst 
Algorithm \ref{alg:MCMC} selected 20 predictors with 17 predictors in common
from the two approaches. This suggests reasonable accuracy
of the faster variational approach for this example.

\section{Concluding Remarks}\label{sec:conclud}

The methodology conveyed by Algorithms \ref{alg:suffStats}--\ref{alg:MFVB} and the effect
type classification rules given in Sections \ref{sec:ZeroVsLin} and \ref{sec:ZeroVsLinVsNonlin}
represent a practical Bayesian alternative to the frequentist methodology 
of Chouldechova \myand Hastie (2015) for three-category generalized additive model selection. 
Both approaches are driven by the goals of speed and scalability to large data sets. 
The new Bayesian approach is the clear winner in terms of accuracy according to 
our simulation studies.

Our, admittedly limited, simulation studies indicate improved  classification performance 
compared with default use of \textsf{spikeSlabGAM} in binary response situations.
In the Gaussian response situations the performance of Algorithm \ref{alg:MCMC}
and \textsf{spikeSlabGAM} is similar, with Algorithm \ref{alg:MFVB} falling behind 
for higher noise situations. This needs to be mitigated against the vastly improved
speed and scalability, as indicated by Table \ref{tab:computTimes}, of this article's  
new methodologies. The best approach in practice depends on data set size
and time demands, with the simulation results of Sections \ref{sec:pracFitSel} 
and \ref{sec:performance} providing some guidance. Additional simulation results are given 
in Section \ref{sec:addSimRes} of the supplement.

\section*{Acknowledgement}

This research was supported by Australian Research Council grant DP180100597.

\section*{Disclosure Statement}

The authors report that there are no competing interests to declare.

\section*{References}

\bib
Albert, J.H. \myand Chib, S. (1993). Bayesian analysis of binary and polychotomous
response data. \textit{Journal of the American Statistical Association},
\textbf{88}, 669--679.

\bib
Azzalini, A. (2023). \textsf{sn 2.1.1}: The Skew-Normal and related distributions
such as the Skew-t and the Unified Skew-Normal. \textsf{R} package.\\
\texttt{http://azzalini.stat.unipd.it/SN}

\bib
Bhadra, A., Datta, J., Polson, N.G. \myand Willard, B. (2019).
Lasso meets horseshoe: a survey. 
\textit{Statistical Science}, \textbf{34}, 405--427.

\bib
B\"urkner, P.-C. (2022).
\textsf{bmrs 2.18.0}: Bayesian regression models using \textsf{Stan}. \textsf{R} package.\\ 
\texttt{https://r-project.org}

\bib
Carvalho, C.M., Polson, N.G. \myand Scott, J.G. (2010).
The horseshoe estimator for sparse signals.
\textit{Biometrika}, \textbf{97}, 465--480.

\bib
Chouldechova, A. \myand Hastie, T. (2015). Generalized additive model selection.\\ 
\texttt{https://arXiv.org/abs/1506.03850v2}

\bib
Chouldechova, A. \myand Hastie, T. (2022).
\textsf{gamsel 1.8}: Fit regularization path for generalized additive models.
\textsf{R} package. \texttt{https://r-project.org}

\bib
Croissant, Y. (2022).
\textsf{Ecdat 0.4}: Data sets for econometrics. \textsf{R} package.\\ 
\texttt{https://r-project.org}

\bib
Eddelbuettel, D. \myand Fran{\c{c}}ois, R. (2011). \textsf{Rcpp}: 
Seamless \textsf{R} and \textsf{C++} integration. 
\textit{Journal of Statistical Software}, \textbf{40(8)}, 1--18.

\bib
Gelfand, A.E. \myand Smith, A.F.M. (1990). Sampling-based approaches
to calculating marginal densities. \textit{Journal of the American Statistical
Association}, \textbf{85}, 398--409.

\bib
Gelman, A. (2006).  Prior distributions for variance parameters 
in hierarchical models. {\it Bayesian Analysis}, {\bf 1}, 515--533.

\bib
George \myand McCulloch, R.E. (1993). Variable selection via
Gibbs sampling. \textit{Journal of the American Statistical Association}, 
\textbf{88}, 881--889.

\bib
Griffin, J.E. \myand Brown, P.J. (2011). Bayesian hyper-lassos with 
non-convex penalization. \textit{Australian and New Zealand Journal
of Statistics}, \textbf{53}, 423--442.

\bib
Harezlak, J., Ruppert, D. \myand Wand, M.P. (2021).
\textsf{HRW 1.0}: Datasets, functions and scripts for semiparametric regression
supporting Harezlak, Ruppert \myand Wand (2018).
\textsf{R} package. \texttt{https://r-project.org}

\bib
Hastie, T.J. \myand Tibshirani, R.J. (1990). \textit{Generalized Additive Models.}
New York: Chapman \myand Hall.

\bib
He, V.X. \myand Wand, M.P. (2023). \textsf{gamselBayes}:
Bayesian generalized additive model selection.
\textsf{R} package version 2.0.\\ 
\texttt{http://cran.r-project.org}.

\bib
Ishwaran, H. \myand Rao, J.S. (2005). Spike and slab variable selection: frequentist
and Bayesian strategies. \textit{The Annals of Statistics}, \textbf{33}, 730--733.

\bib
Kyung, M., Gill, J., Ghosh, M. \myand Casella, G. (2010).
Penalized regression, standard errors, and Bayesian lassos. 
\textit{Bayesian Analysis}, \textbf{5}, 369--412.

\bib
Lempers, F.B. (1971). \textit{Posterior Probabilities of Alternative Linear Models.}
Rotterdam: Rotterdam University Press.

\bib
Michael, J.R., Schucany, W.R. \myand Haas, R.W. (1976).
Generating random variates using transformations with multiple roots.
\textit{The American Statistician}, \textbf{30}, 88--90.

\bib
Mitchell, T.J. \myand Beauchamp, J.J. (1988).
Bayesian variable selection in linear regression.
\textit{Journal of the American Statistical Association},
\textbf{83}, 1023--1032.

\bib
Ngo, L. and Wand, M.P. (2004).
Smoothing with mixed model software.
{\it Journal of Statistical Software},
{\bf 9}, Article 1, 1--54.

\bib
Ormerod, J.T. and Wand, M.P. (2010).
Explaining variational approximations.
{\it The American Statistician},
{\bf 64}, 140--153.

\bib
Park, T. \myand Casella, G. (2008). The Bayesian Lasso.
\textit{Journal of the American Statistical Association}, 
\textbf{103}, 681--686.

\bib
Ravikumar, P., Lafferty, J.,  Liu, H. \myand Wasserman, L. (2009). 
Sparse additive models.\ {\it Journal of the Royal Statistical
Society, Series B}, {\bf 71}, 1009--1030.

\bib
\textsf{R} Core Team (2023). \textsf{R}: A language and environment
for statistical computing. \textsf{R} Foundation for Statistical
Computing, Vienna, Austria. \texttt{https://www.r-project/org/}.

\bib
Reich, B.J., Sorlie, C.B. \myand Bondell, H.D. (2009). 
Variable selection in smoothing spline ANOVA: application to
deterministic computer codes.\ {\it Technometrics}, 
{\bf 51}, 110--120.

\bib
Robert, C.P. (1995). Simulation of truncated normal variates.
\textit{Statistics and Computing}, \textbf{5}, 121--125.

\bib
Scheipl, F. (2011). \textsf{spikeSlabGAM}: Bayesian variable selection, model
choice and regularization for generalized additive mixed models in \textsf{R}.
\textit{Journal of Statistical Software}, \textbf{43}, Issue 14. 1--24.

\bib
Scheipl, F. (2022). \textsf{spikeSlabGAM 1.1}: Bayesian variable selection and
model choice for generalized additive mixed models.
\textsf{R} package.\\ 
\texttt{https://github.com/fabian-s/spikeSlabGAM}

\bib
Scheipl, F., Fahrmeir, L. \myand Kneib, T. (2012). 
Spike-and-slab priors for function selection in structured
additive regression models.\ {\it Journal of the American
Statistical Association}, {\bf 107}, 1518--1532.

\bib
Shively, T.S., Kohn, R. \myand Wood, S. (1999). 
Variable selection and function estimation in additive nonparametric
regression using a data-based prior.\ {\it Journal of the American
Statistical Association}, {\bf 94}, 777--794.

\bib
Umlauf, N., Klein, N., Zeileis, A. \myand Simon, T. (2023).
\textsf{bamlss 1.2}: Bayesian additive models for location, scale, and shape (and beyond).
\textsf{R} package. \texttt{https://www.bamlss.org}

\bib
Umlauf, N., Kneib, T. \myand Klein, N. (2023).
\textsf{BayesX 0.3}: \textsf{R} utilities accompanying the software package \textsf{BayesX}.
\textsf{R} package. \texttt{https://www.BayesX.org}

\bib
Wainwright, M.J. \myand Jordan, M.I. (2008).
Graphical models, exponential families and variational 
inference. \textit{Foundations and Trends in 
Machine Learning}, {\bf 1}, 1--305.

\bib
Wand, M.P. \myand Ormerod, J.T. (2008).
On semiparametric regression with O'Sullivan penalized splines.
{\it Australian and New Zealand Journal of Statistics},
{\bf 50}, 179--198.

\bib
Wand, M.P. and Ormerod, J.T. (2011).
Penalized wavelets: embedding wavelets into semiparametric regression.
{\it Electronic Journal of Statistics},
{\bf 5}, 1654--1717.

\bib
Wand, M.P. and Ormerod, J.T. (2012).
Continued fraction enhancement of Bayesian computing.
\textit{Stat}, {\bf 1}, 31--41.

\bib
Wood, S.N. (2017). \textit{Generalized Additive Models: An Introduction with
\textsf{R}, Second Edition}, Boca Raton, Florida: CRC Press.

\bib
Yuan, M. \myand Lin, Y. (2006). Model selection and estimation in
regression with grouped variables. \textit{Journal of the Royal
Statistical Society, Series B}, \textbf{68}, 49--67.


\null\vfill\eject

%
%


\renewcommand{\theequation}{S.\arabic{equation}}
\renewcommand{\thesection}{S.\arabic{section}}
\renewcommand{\thetable}{S.\arabic{table}}
\renewcommand{\thefigure}{S.\arabic{figure}}
\setcounter{equation}{0}
\setcounter{table}{0}
\setcounter{figure}{0}
\setcounter{section}{0}
\setcounter{page}{1}
\setcounter{footnote}{0}
\begin{center}
{\Large Supplement for:}
\vskip3mm

\centerline{\Large\bf Bayesian Generalized Additive Model Selection}
\vskip1mm
\centerline{\Large\bf Including a Fast Variational Option}
\vskip5mm
\centerline{\normalsize\sc By Virginia X. He and Matt P. Wand}
\vskip5mm
\centerline{\textit{University of Technology Sydney}}
\end{center}

\section{The Canonical Demmler-Reinsch Spline Basis}\label{sec:ZcDR}

Let $\bx=(x_1,\ldots,x_n)$ be a continuous univariate data set. In the context of this
article, the $x_i$s correspond to values of a continuous candidate predictor.
Let $[a,b]$ be an interval containing the $x_i$s. For an integer $K\le n-2$,
let $\bkappaInterior\equiv(\kappa_1,\ldots,\kappa_{K-2})$ be a set of so-called interior 
knots such that
$$a<\kappa_1<\cdots<\kappa_{K-2}<b.$$
A reasonable default value for $K$ is around $30$, or smaller values if the number
of unique $x_i$s is lower. It is common to place the interior knots at sample
quantiles of the $x_i$s.

We now list steps for construction of the matrix $\bZ$ containing canonical 
Demmler-Reinsch basis functions of the entries of $\bx$. The justification for
Steps (3)--(6) is given in Section 9.1.1 of Ngo \myand Wand (2004).
\begin{itemize}
\item[(1)] Use the steps described in Section 4 of Wand \myand Ormerod (2008) to obtain the 
matrix denoted by $\bZ$ in that section's equation (6), which contains canonical O'Sullivan 
spline basis functions. Denote this matrix by $\bZOS$ and note that it has dimension $n\times K$.
\item[(2)] Form the matrix $\bCOS=[\bone_n\ \bx\ \bZOS]$ 
and set $\bD=\diag(0,0,\bone_K)$.
\item[(3)] Obtain the singular value decomposition of $\bCOS$:
\begin{eqnarray*}
&&\bCOS=\bUC\diag(\bdC)\bVC^T\ \mbox{where $\bUC$ is $n\times(K+2)$ and $\bVC$ is $(K+2)\times(K+2)$}\\[1ex] 
&&\mbox{such that}\ \bUC^T\bUC=\bVC^T\bVC=\bI_{K+2}.
\end{eqnarray*}
\item[(4)] Form the symmetric matrix $\diag(\bone/\bdC)\bVC^T\bD\bVC\diag(\bone/\bdC)$
and obtain its singular value decomposition:
\begin{eqnarray*}
&&\diag(\bone/\bdC)\bVC^T\bD\bVC\diag(\bone/\bdC)=\bUD\diag(\bdD)\bVD^T
\ \mbox{where $\bUD$ is $(K+2)\times(K+2)$}\\[1ex] 
&&\mbox{and $\bVD$ is $(K+2)\times(K+2)$ such that}\ \bUD^T\bUD=\bVD^T\bVD=\bI_{K+2}.
\end{eqnarray*}
\item[(5)]  Set the full (non-canonical)
Demmler-Reinsch matrix as follows: $\bCDR\thickarrow\bUC\bUD$.
\item[(6)] The next steps assume that the singular value decompositions follow the 
convention that $\bdD$ is a $(K+2)\times1$ vector with its entries in non-increasing order.
Adjustments to the singular value decompositions are needed if this convention is not
used.
\item[(7)] Set the $(K+2)\times1$ vector $\bsD$ as follows:
$$\omega_{21}\thickarrow\sqrt{\mbox{$K$th entry of $\bdD$}},\ \ \ ;\ \ \ 
\bsD\thickarrow\omega_{21}\bone_{K+2}\Big/\sqrt{\bdD}$$
and then set the last two entries of $\bsD$ to equal $1$.
\item[(8)] Set the full canonical Demmler-Reinsch design matrix as follows:
$$\bCcDR\thickarrow\bCDR\diag(\bsD).$$
\item[(9)] Set the O'Sullivan to canonical Demmler-Reinsch transformation matrix as follows:
$$\bLOStocDR\thickarrow \bVC\diag(\bone/\bdC)\bUD\diag(\bsD).$$
This $(K+2)\times(K+2)$ matrix has the following property:
$$\bCOS\bLOStocDR=\bCcDR$$
and is useful for prediction and plotting purposes. This is because grid-wise analogues of $\bCOS$ 
are readily computed using the structures described in Wand \myand Ormerod (2008) involving cubic 
B-spline basis functions.
\item[(10)] Reverse the order of the columns of $\bCcDR$. Reverse the order of the columns of $\bLOStocDR$.
\item[(11)] The matrix containing canonical spline basis functions of the inputs $\bx$ 
and $\bkappaInterior$ is 
$$\bZ\thickarrow\mbox{the $n\times K$ matrix consisting of columns $3$ to $K+2$ of $\bCcDR$.}$$
\end{itemize}

A function in the \textsf{R} language for computing $\bZ$ and $\bLOStocDR$
for given $\bx$ and $\bkappaInterior$ can be accessed by downloading the accompanying
\textsf{gamselBayes} package. Assuming that the \textsf{gamselBayes} package is
installed, the relevant function is \texttt{gamselBayes:::ZcDR()}.

\section{Approximate Marginal Log-Likelihood Expressions}\label{sec:logML}

The approximate marginal log-likelihood is
$$
\log\punder(\by;\qDens)=
\left\{
\begin{array}{lcl}
\log\punder(\by;\qDens,\COMMONabv)+E_{\qDens}
[\log\{\pDens(\by|\beta_0,\gammaSUBbeta,\bbetatilde,\gammaSUBu,\butilde,\sigeps^2)\}]\\[1ex]
\quad+E_{\qDens}[\log\{\pDens(\sigsqeps|\aSUBeps)\}]
-E_{\qDens}[\log\{\qDens(\sigsqeps)\}]\\[1ex]
\quad+E_{\qDens}[\log\{\pDens(\aSUBeps)\}]-E_{\qDens}[\log\{\qDens(\aSUBeps)\}]
&\mbox{Gaussian response case,}\\[2ex]
\log\punder(\by;\qDens,\COMMONabv)
+E_{\qDens}[\log\{\pDens(\by|\bc)\}]\\[1ex]
\quad+E_{\qDens}
[\log\{\pDens(\bc|\beta_0,\gammaSUBbeta,\bbetatilde,\gammaSUBu,\butilde)\}]
-E_{\qDens}[\log\{\qDens(\bc)\}] &\mbox{Bernoulli response case,}
\end{array}
\right.
$$
where
\begin{equation}
{\setlength\arraycolsep{0pt}
\begin{array}{rcl}
&&\log\punder(\by;\qDens,\COMMONabv)=
E_{\qDens}[\log\{\pDens(\beta_0)\}]-E_{\qDens}[\log\{\qDens(\beta_0)\}]
+E_{\qDens}[\log\{\pDens(\gammaSUBbeta)\}]
-E_{\qDens}[\log\{\qDens(\gammaSUBbeta)\}]
\\[1ex]
&&\qquad +E_{\qDens}[\log\{\pDens(\bbetatilde|\bbSUBbeta,\sigsqSUBbeta)\}]
-E_{\qDens}[\log\{\qDens(\bbetatilde)\}]
+E_{\qDens}[\log\{\pDens(\bbSUBbeta)\}]-E_{\qDens}[\log\{\qDens(\bbSUBbeta)\}]
\\[1ex]
&&\qquad +E_{\qDens}[\log\{\pDens(\sigsqSUBbeta|\aSUBbeta)\}]
-E_{\qDens}[\log\{\qDens(\sigsqSUBbeta)\}]
+E_{\qDens}[\log\{\pDens(\aSUBbeta)\}]-E_{\qDens}[\log\{\qDens(\aSUBbeta)\}]\\[1ex]
&&\qquad +E_{\qDens}[\log\{\pDens(\gammaSUBu)\}]
-E_{\qDens}[\log\{\qDens(\gammaSUBu)\}]
+E_{\qDens}[\log\{\pDens(\butilde|\bbSUBu,\bsigsqSUBu)\}]
-E_{\qDens}[\log\{\qDens(\butilde)\}]
\\[1ex]
&&\qquad 
+E_{\qDens}[\log\{\pDens(\bbSUBu)\}]-E_{\qDens}[\log\{\qDens(\bbSUBu)\}]
+E_{\qDens}[\log\{\pDens(\bsigsqSUBu|\baSUBu)\}]
-E_{\qDens}[\log\{\qDens(\bsigsqSUBu)\}]
\\[1ex]
&&\qquad 
+E_{\qDens}[\log\{\pDens(\baSUBu)\}]
-E_{\qDens}[\log\{\qDens(\baSUBu)\}].
\end{array}
}
\label{eq:logMLexpanCOMMON}
\end{equation}
Here ``\COMMONabv'' signifies the fact that (\ref{eq:logMLexpanCOMMON}) is  
common to both $\log\punder(\by;\qDens)$ expressions.

Explicit expressions for $\log\punder(\by;\qDens)$ can be
obtained by simplifying each of the $\qDens$-density moment expressions.
For example, the first term of (\ref{eq:logMLexpanCOMMON}) is
{\setlength\arraycolsep{1pt}
\begin{eqnarray*}
E_{\qDens}[\log\{\pDens(\beta_0)\}]&=&-\smhalf\log(2\pi)-\smhalf\log(\sigma_{\beta_0}^2)
-\smhalf E_{\qDens}(\beta_0^2)\big/\sigma_{\beta_0}^2\\[1ex]
&=&-\smhalf\log(2\pi)-\smhalf\log(\sigma_{\beta_0}^2)
-\smhalf\{\mu^2_{\qDens(\beta_0)}+\sigma^2_{\qDens(\beta_0)}\}\big/\sigma_{\beta_0}^2.
\end{eqnarray*}
}
Also, since $\qDens(\beta_0)$ is the $N\big(\mu_{\qDens(\beta_0)},\sigma_{\qDens(\beta_0)}^2\big)$
density function, the second term of (\ref{eq:logMLexpanCOMMON}) is
{\setlength\arraycolsep{1pt}
\begin{eqnarray*}
-E_{\qDens}[\log\{\qDens(\beta_0)\}]&=&\smhalf\log(2\pi)
+\smhalf\log\big(\sigma_{\qDens(\beta_0)}^2\big)
+\smhalf 
E_{\qDens}\big\{\big(\beta_0-\mu_{\qDens(\beta_0)}\big)^2\big\}\big/\sigma_{\qDens(\beta_0)}^2\\[1ex]
&=&\smhalf\{\log(2\pi)+1\}+\smhalf\log\big(\sigma_{\qDens(\beta_0)}^2\big).
\end{eqnarray*}
}
Continuing in this fashion, and accounting for some cancellations, we obtain
{\setlength\arraycolsep{1pt}
\begin{eqnarray*}
\log\punder(\by;\qDens,\COMMONabv)&=&\mbox{const}_1
-\smhalf\{\mu^2_{\qDens(\beta_0)}+\sigma^2_{\qDens(\beta_0)}\}\big/\sigma_{\beta_0}^2
+\smhalf\log\big(\sigma_{\qDens(\beta_0)}^2\big)
+\logit(\rhoSUBbeta) \sum_{j=1}^{\dLin+\dNon}\mu_{\qDens(\gammaSUBbetaj)}
\\[1ex]
&&\quad-\sum_{j=1}^{\dLin+\dNon}
\left[\mu_{\qDens(\gammaSUBbetaj)}\log\big(\mu_{\qDens(\gammaSUBbetaj)}\big)
+\{1-\mu_{\qDens(\gammaSUBbetaj)}\}
\log\big(1-\mu_{\qDens(\gammaSUBbetaj)}\big)\right]\\[1ex]
&&\quad -\smhalf\mu_{\qDens(1/\sigsqSUBbeta)}
\sum_{j=1}^{\dLin+\dNon}\mu_{\qDens(\bSUBbetaj)}\big(\mu_{\qDens(\betatilde_j)}^2
+\sigma_{\qDens(\betatilde_j)}^2\big)+\smhalf\log\big|\bSigma_{\qDens(\bbetatilde)}\big|
\\[1ex]
&&\quad -\smhalf\sum_{j=1}^{\dLin+\dNon}\{1/\mu_{\qDens(\bSUBbetaj)}\}
-\mu_{\qDens(1/\aSUBbeta)}\mu_{\qDens(1/\sigsqSUBbeta)}
-\smhalf(\dLin+\dNon+1)\log\big(\lambda_{\qDens(\sigsqSUBbeta)}\big)
\\[1ex]
&&\quad +\mu_{\qDens(1/\sigsqSUBbeta)}\lambda_{\qDens(\sigsqSUBbeta)}-\mu_{\qDens(1/\aSUBbeta)}/\sSUBbeta^2
+\lambda_{\qDens(\aSUBbeta)}\mu_{\qDens(1/\aSUBbeta)}-\log\big(\lambda_{\qDens(\aSUBbeta)})
\\[1ex]
&&\quad -\sum_{j=1}^{\dNon}
\left[\mu_{\qDens(\gammaSUBuj)}\log\big(\mu_{\qDens(\gammaSUBuj)}\big)
+\{1-\mu_{\qDens(\gammaSUBuj)}\}
\log\big(1-\mu_{\qDens(\gammaSUBuj)}\big)\right]
\\[1ex]
&&\quad +\logit(\rhoSUBu) \sum_{j=1}^{\dNon}\mu_{\qDens(\gamma_{uj})}
-\smhalf\sum_{j=1}^{\dNon}\mu_{\qDens(1/\sigsqSUBuj)}\mu_{\qDens(\bSUBuj)}
\Big(\Vert\bmu_{\qDens(\butilde_j)}\Vert^2+\bone_{K_j}^T\bsigma^2_{\qDens(\butilde_j)}\Big)
\\[1ex]
&&\quad +\smhalf\sum_{j=1}^{\dNon}\sum_{k=1}^{K_j}\log\big(\sigma^2_{\qDens(\butilde_{jk})}\big)
-\smhalf\sum_{j=1}^{\dNon}\{1/\mu_{\qDens(\bSUBuj)}\}
-\sum_{j=1}^{\dNon}\mu_{\qDens(1/\aSUBuj)}\mu_{\qDens(1/\sigsqSUBuj)}
\\[1ex]
&&\quad
-\smhalf\sum_{j=1}^{\dNon}(K_j+1)\log\big(\lambda_{\qDens(\sigsqSUBuj)}\big)
+\sum_{j=1}^{\dNon}\mu_{\qDens(1/\sigsqSUBuj)}\lambda_{\qDens(\sigsqSUBuj)}
-(1/s_u^2)\sum_{j=1}^{\dNon}\mu_{\qDens(1/\aSUBuj)}\\[1ex]
&&\quad
+\sum_{j=1}^{\dNon}\big\{\lambda_{\qDens(\aSUBuj)}\mu_{\qDens(1/\aSUBuj)}
-\log\big(\lambda_{\qDens(\aSUBuj)}\big)\big\}
\end{eqnarray*}
}
where $\mbox{const}_1$ is a constant that does not depend on any $\qDens$-density parameters. 

In the Gaussian response case, we have
{\setlength\arraycolsep{1pt}
\begin{eqnarray*}
\log\punder(\by;\qDens)&=&
\log\punder(\by;\qDens,\COMMONabv)
-\smhalf(n+1)\log(\lambda_{\qDens(\sigsqSUBeps)})
-\mu_{\qDens(1/\aSUBeps)}/\sSUBeps^2
-\log\big(\lambda_{\qDens(\aSUBeps)}\big)+\lambda_{\qDens(\aSUBeps)}\mu_{\qDens(1/\aSUBeps)}\\[1ex]
&&\qquad+\mbox{const}_2,
\end{eqnarray*}
}
where $\mbox{const}_2$ is a constant that does not depend on any $\qDens$-density
parameters. In the Bernoulli response case
{\setlength\arraycolsep{1pt}
\begin{eqnarray*}
\log\punder(\by;\qDens)&=&
\log\punder(\by;\qDens,\COMMONabv)
+\displaystyle{\sumin} \log\Big\{\Phi\Big((2y_i-1)\Big(\bone_n\mu_{\qDens(\beta_0)}
+\bX\big(\bmu_{\qDens(\gammaSUBbeta)}\odot\bmu_{\qDens(\bbetatilde)}\big)\\[0ex]
&&\qquad\qquad\qquad\qquad\qquad+{\displaystyle\sum_{j=1}^{\dNon}}\bZ_j
\big(\mu_{\qDens(\gammaSUBuj)}\bmu_{\qDens(\butilde_j)}\big)\Big)_i\Big)\Big\}.
\end{eqnarray*}
}

\section{Additional Simulation Results}\label{sec:addSimRes}

We have conducted thorough simulation testing of Algorithms \ref{alg:MCMC} and \ref{alg:MFVB}
and the model selection strategies given in Section \ref{sec:modSelecStrat}.
Space considerations are such that Sections \ref{sec:pracFitSel} and \ref{sec:performance} 
contain only our primary simulation results. Additional simulation results are conveyed
in this section.

\begin{figure}[h]
\centering
\includegraphics[width=1.0\textwidth]{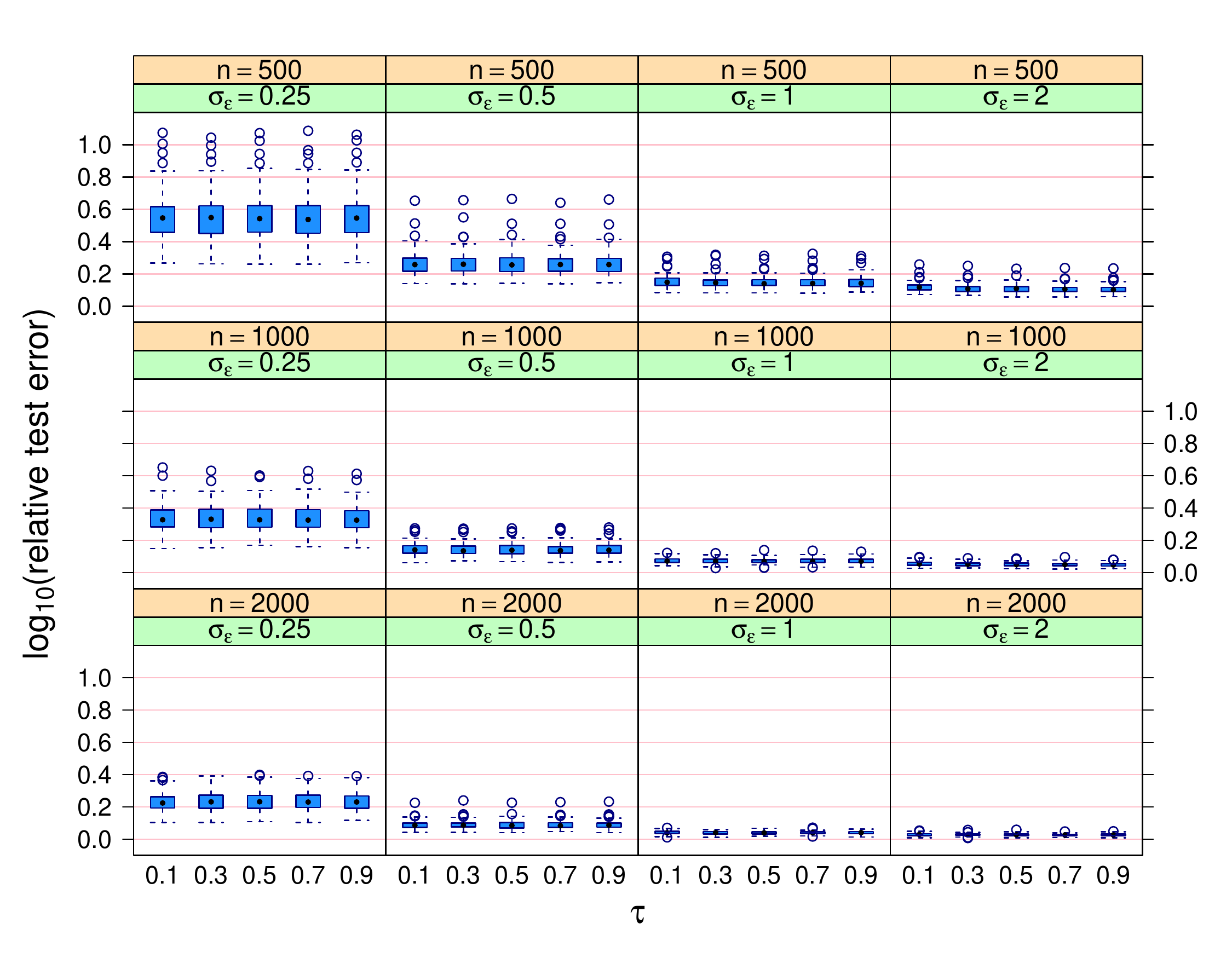}
\caption{
\textit{Side-by-side boxplots of the logarithms, to base $10$,
of relative test error for the Markov chain Monte Carlo Algorithm \ref{alg:MCMC}
for the simulation study described in the text. Each panel corresponds
to a different combination of sample size and error standard deviation.
Within each panel, the side-by-side boxplots compare relative test error
as a function of the threshold parameter $\tau$.}}
\label{fig:choiceTauMCMCgaussRTE}
\end{figure}

\subsection{Alternative Evaluation Metrics}

The model selection recommendations of Section \ref{sec:modSelecStrat} are
guided by the effect type misclassification rate since we believe this
particular evaluation metric to be best aligned with the practical goal of achieving 
interpretable and parsimonious models. However effect type misclassification
rate is just one of many possible evaluation metrics that could be used
in simulation assessment, comparison and the guiding of tuning parameter choice.
For example, the simulation studies of Hastie, Tibshirani \myand Tibshirani (2020),
for a different regression-type setting, consider five evaluation metrics.

To see if and how our threshold parameter recommendations change if a different
evaluation metric is used, we re-ran the Gaussian response simulation studies of 
Section \ref{sec:modSelecStrat} with effect type misclassification rate replaced 
by \emph{relative test error}. For the situation where $\dLin=0$ and $\dNon\in\naturalNumbers$,
suppose that the selected model based on the data set $\bDscr$ corresponds to $\fhat$ 
for some additive function $\fhat:\real^{\dNon}\to\real$. If the true model 
corresponds to $\fTrue:\real^{\dNon}\to\real$ and the predictor $\bx\in\real^{\dNon}$
is a random vector with density function $\pDens(\bx)$ then the relative test error is 
\begin{equation}
E\big[\big\{y-\fhat(\bx)\}^2|\bDscr\big]\big/\sigeps^2\quad\mbox{where}\quad 
y\sim N\big(\fTrue(\bx),\sigeps^2\big).
\label{eq:RTEdefn}
\end{equation}
Note that the expectation in (\ref{eq:RTEdefn}) is over the predictor distribution corresponding
to $\pDens(\bx)$. The denominator in (\ref{eq:RTEdefn}) is the \emph{Bayes error},
corresponding to the situation where $\fhat=\fTrue$. Therefore, (\ref{eq:RTEdefn}) is the
test error relative to the Bayes error and is an evaluation metric with a lower bound of $1$,
and equals $1$ when $\fTrue$ is estimated perfectly.

\begin{figure}[h]
\centering
\includegraphics[width=1.0\textwidth]{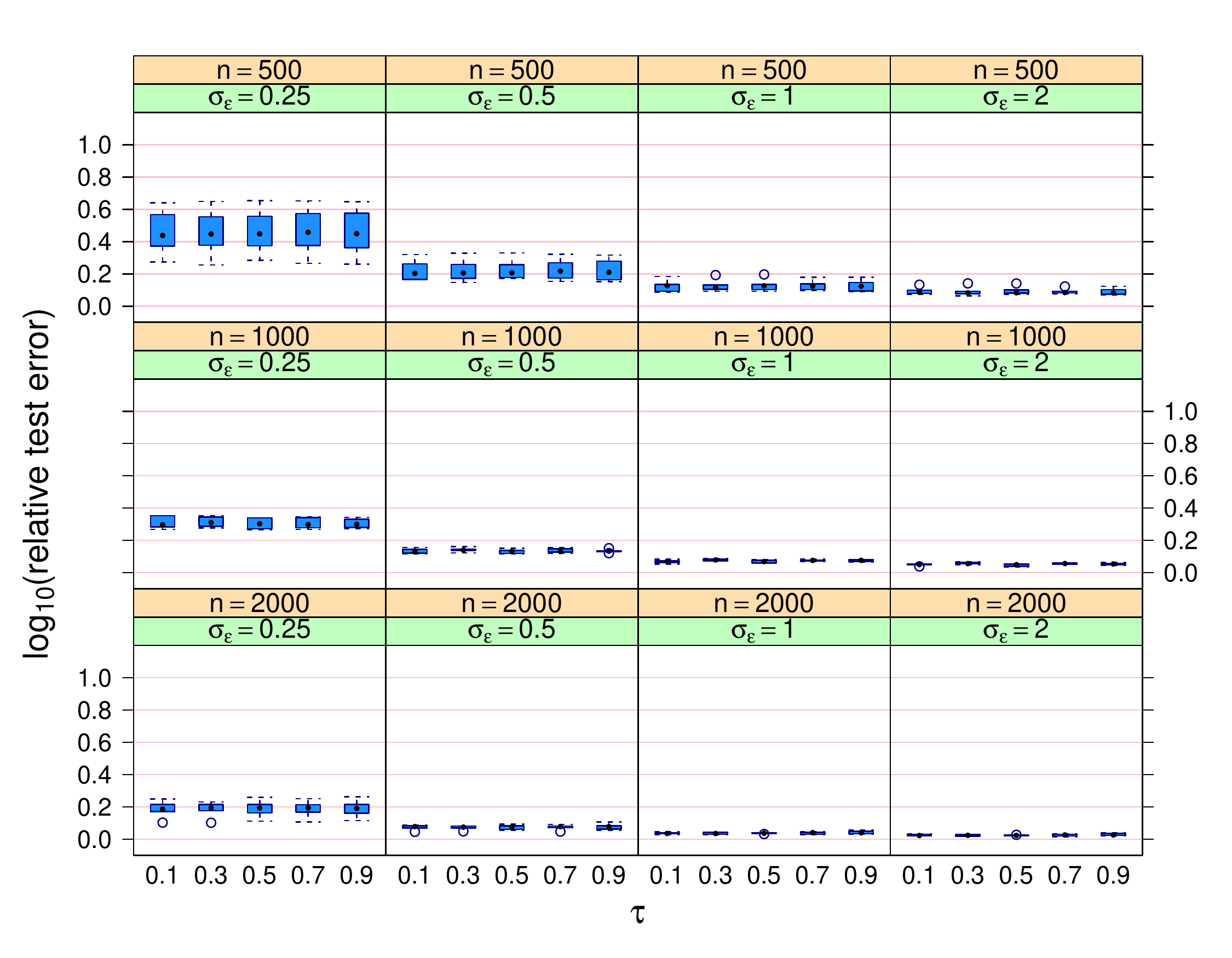}
\caption{
\textit{Side-by-side boxplots of the logarithms, to base $10$,
of relative test error for the mean field variational Bayes Algorithm \ref{alg:MFVB}
for the simulation study described in the text. Each panel corresponds
to a different combination of sample size and error standard deviation.
Within each panel, the side-by-side boxplots compare relative test error
as a function of the threshold parameter $\tau$.}}
\label{fig:choiceTauMFVBgaussRTE}
\end{figure}

Figures \ref{fig:choiceTauMCMCgaussRTE} and \ref{fig:choiceTauMFVBgaussRTE}
are the analogues of Figures \ref{fig:choiceTauMCMCgauss} and \ref{fig:choiceTauMFVBgauss}
with effect type misclassification rate replaced by relative test error. Monte Carlo
approximations of the (\ref{eq:RTEdefn}) numerator quantity based on 100,000 draws
from the predictor distribution were used. To aid visualization the $\log_{10}$ transformation
is applied to the relative test error values.

From Figures \ref{fig:choiceTauMCMCgaussRTE} and \ref{fig:choiceTauMFVBgaussRTE}
we see that the relative test errors are lower for higher sample sizes, as expected. 
Somewhat counter-intuitively the relative test errors are lower for
higher noise levels. However, comparisons of relative test error
across different values of $\sigeps$ are not clear-cut when the
estimators are subject to bias.
In addition, relative test errors are barely affected by the choice of the 
thresholding parameter $\tau$.
Lastly, the relative test errors based on mean field variational Bayes 
approximate inference are similar to those based on Markov chain
Monte Carlo. It is interesting that this particular evaluation metric
is not affected very much by the choices between Algorithms \ref{alg:MCMC} and
\ref{alg:MFVB} and the value of the threshold parameter $\tau$.

\subsection{Detailed Computing Time Results}\label{sec:detCompTimeRes}

We also conducted some more detailed involving computing times.
One simulation study looked into the effect of sample size, whilst
another one investigated how the number of candidate predictors
impacts computing times. The results are presented in this section.

\subsubsection{Assessment of the Effect of Sample Size}

Our first detailed computing time simulation study was concerned with 
the effect of sample size. We fixed the candidate predictor dimensions
to be $(\dLin,\dNon)=(0,10)$ and let the sample size $n$ to range over
the set
$$\{10^k:k=2,3,4,5,6\}.$$
The data were generated in a manner similar to that for the simulation
studies described in Sections \ref{sec:pracFitSel} and \ref{sec:performance},
with $100$ replications.

\begin{figure}[h]
\centering
\includegraphics[width=1.0\textwidth]{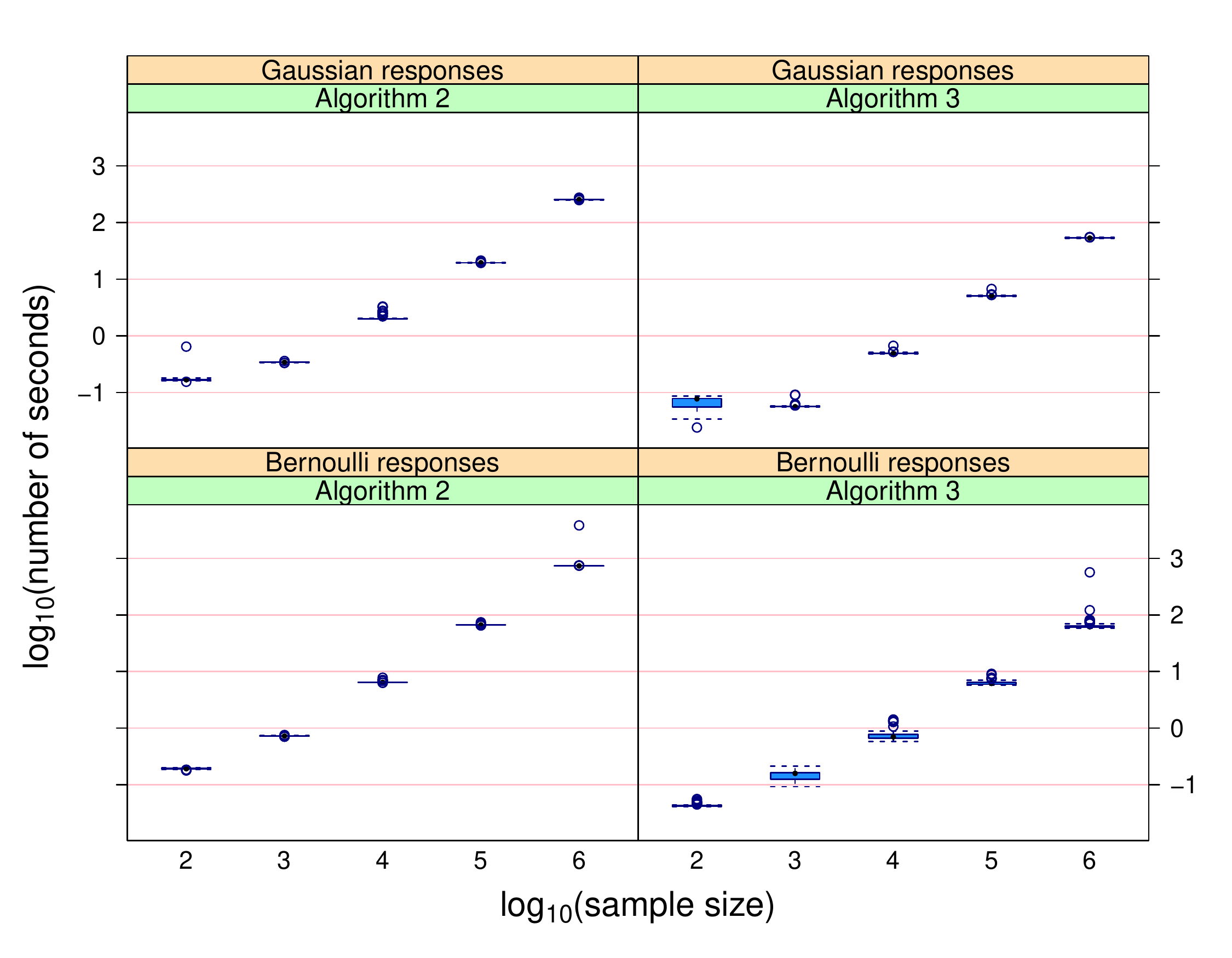}
\caption{
\textit{Side-by-side box plots of computing time in seconds
versus sample size for generalized additive
model selection via Algorithms \ref{alg:MCMC} and \ref{alg:MFVB},
for the first simulation study described in Section \ref{sec:detCompTimeRes}.
Each axis uses a $\log_{10}$ scale.}}
\label{fig:timeVsSampSizeCol}
\end{figure}

Figure \ref{fig:timeVsSampSizeCol} summarizes the results using
side-by-side boxplots of the logarithmically transformed computing
times, broken down according to sample size, response type
and whether or not Algorithm \ref{alg:MCMC} or Algorithm \ref{alg:MFVB}
was used. The relationships between the mean logarithmic number
of seconds and logarithmic sample size are approximately linear,
which suggests a simple power law relationship between computing 
time and sample size. Simple linear regression analyses suggest
that the power is close to $1$ and, hence, mean computing
time is roughly proportional to sample size.
 
Figure \ref{fig:timeVsSampSizeCol} also shows that use of 
Algorithm \ref{alg:MFVB} leads to an approximately ten-fold
reduction in computing time compared with Algorithm \ref{alg:MCMC}.
For example, when $n=100,000$ the mean computing time of Algorithm 
\ref{alg:MCMC} for Bernoulli responses is about $100$ seconds.
For Algorithm \ref{alg:MFVB} it is only about $10$ seconds.

\subsubsection{Assessment of the Effect of the Number of Candidate Predictors}

We also ran a simulation study concerned with the effect of the number of
candidate predictors on computing time. The sample size was fixed at
$5,000$ and $\dNon$, the number of candidate predictors that could enter the
model non-linearly, varied over the set
$$\{2^k:k=1,2,3,4,5,6\}.$$
We generated the data in a manner similar to that for the simulation
studies described in Sections \ref{sec:pracFitSel} and \ref{sec:performance}
and, again, obtained $100$ replications.

\begin{figure}[h]
\centering
\includegraphics[width=1.0\textwidth]{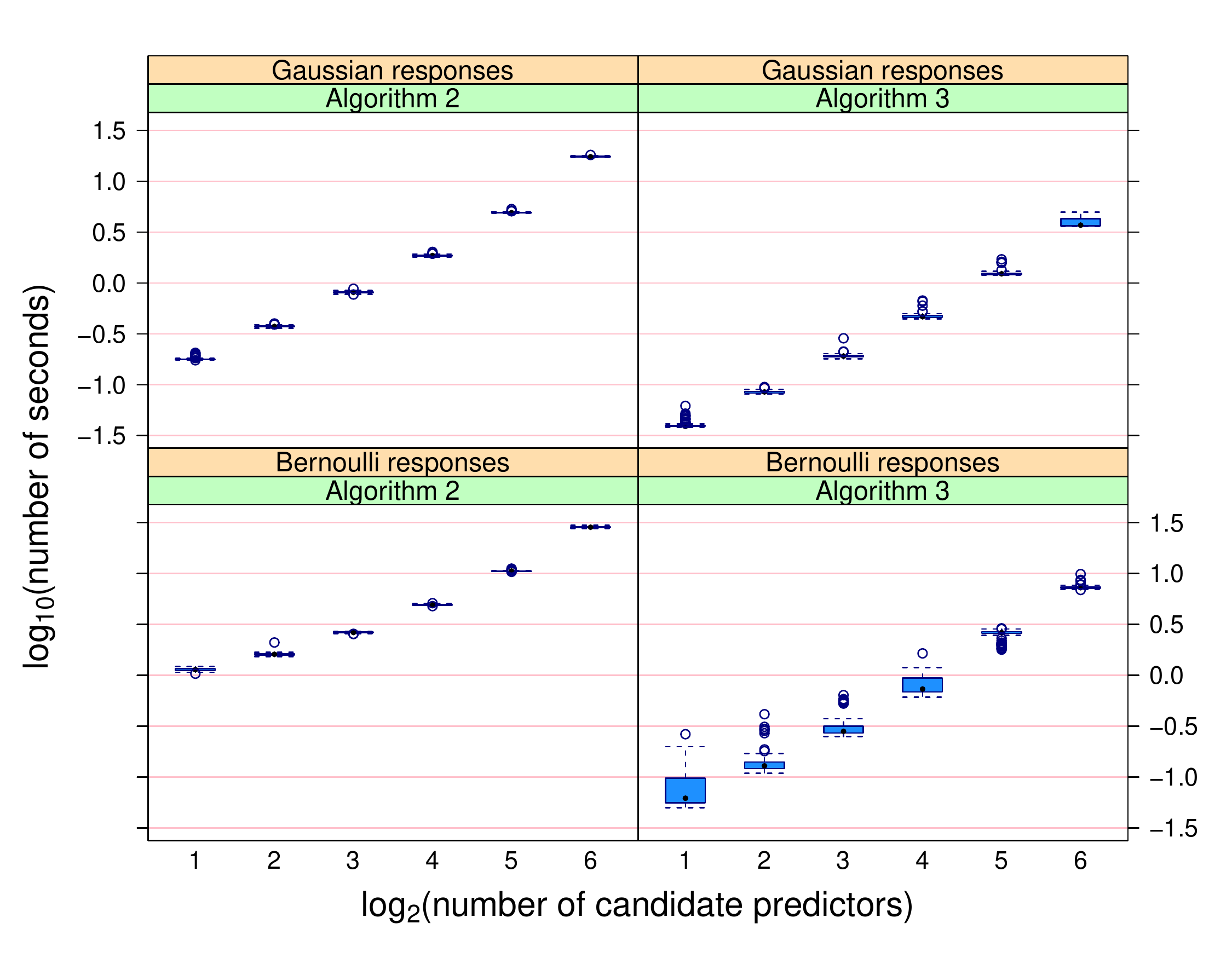}
\caption{
\textit{Side-by-side box plots of computing time in seconds
versus number of candidate predictors for generalized additive
model selection via Algorithms \ref{alg:MCMC} and \ref{alg:MFVB},
for the second simulation study described in Section \ref{sec:detCompTimeRes}.
The horizontal axis uses a $\log_2$ scale and the vertical axis uses
a $\log_{10}$ scale.}}
\label{fig:timeVsNumPredsCol}
\end{figure}

Figure \ref{fig:timeVsNumPredsCol} summarises the results in similar way
to Figure \ref{fig:timeVsSampSizeCol}. Once again, there is approximate
linearity within each panel with logarithmic scales. Simple linear
regression analyses of the data within each panel of 
Figure \ref{fig:timeVsNumPredsCol} suggest that the mean 
computing time is approximately proportional to $\dNon^{\kappa}$,
with $\kappa$ dependent on the response distribution and fitting 
algorithm combination but within the interval $(1.2,1.5)$.

\subsection{Hyperparameter Sensitivity Checks}\label{sec:HypSensChks}

\begin{figure}[h]
\centering
\includegraphics[width=1.0\textwidth]{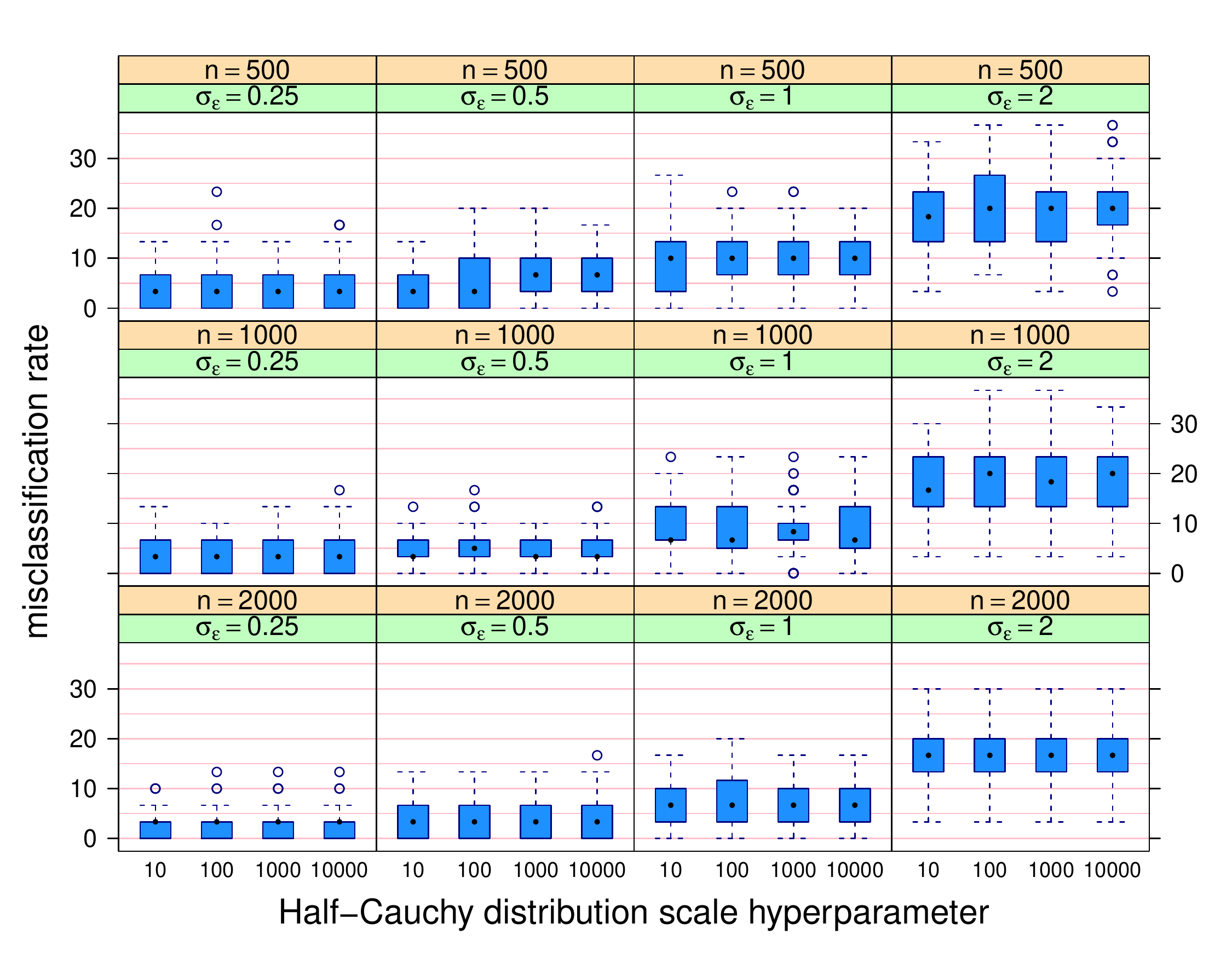}
\caption{
\textit{Side-by-side boxplots of misclassification rate for varying
values of the Half Cauchy distribution scale hyperparameter for the Gaussian response
version of Algorithm \ref{alg:MCMC}, for the first simulation study
described in Section \ref{sec:HypSensChks}.}}
\label{fig:choiceScaHalfCauMCMCgaussCol}
\end{figure}

Figure \ref{fig:choiceScaHalfCauMCMCgaussCol} conveys the effect of the
Half Cauchy distribution scale hyperparameters, denoted by 
$\sSUBbeta$, $\sSUBeps$ and $s_u$ in model (\ref{eq:mainModel}), on
the effect type misclassification rate. It is based on the simulation
study set-up of Section \ref{sec:modSelecStrat} with 
the Markov chain Monte Carlo approach of Algorithm \ref{alg:MCMC}
and the threshold parameter $\tau$ set to our recommended default value of $0.5$. 
The scale hyperparameters ranged over the set
$$\{10^k:k=1,2,3,4\}.$$
Figure \ref{fig:choiceScaHalfCauMCMCgaussCol} indicates that
our default version of Algorithm \ref{alg:MCMC} is not sensitive
to the Half Cauchy distribution scale hyperparameter values.

\begin{figure}[h]
\centering
\includegraphics[width=1.0\textwidth]{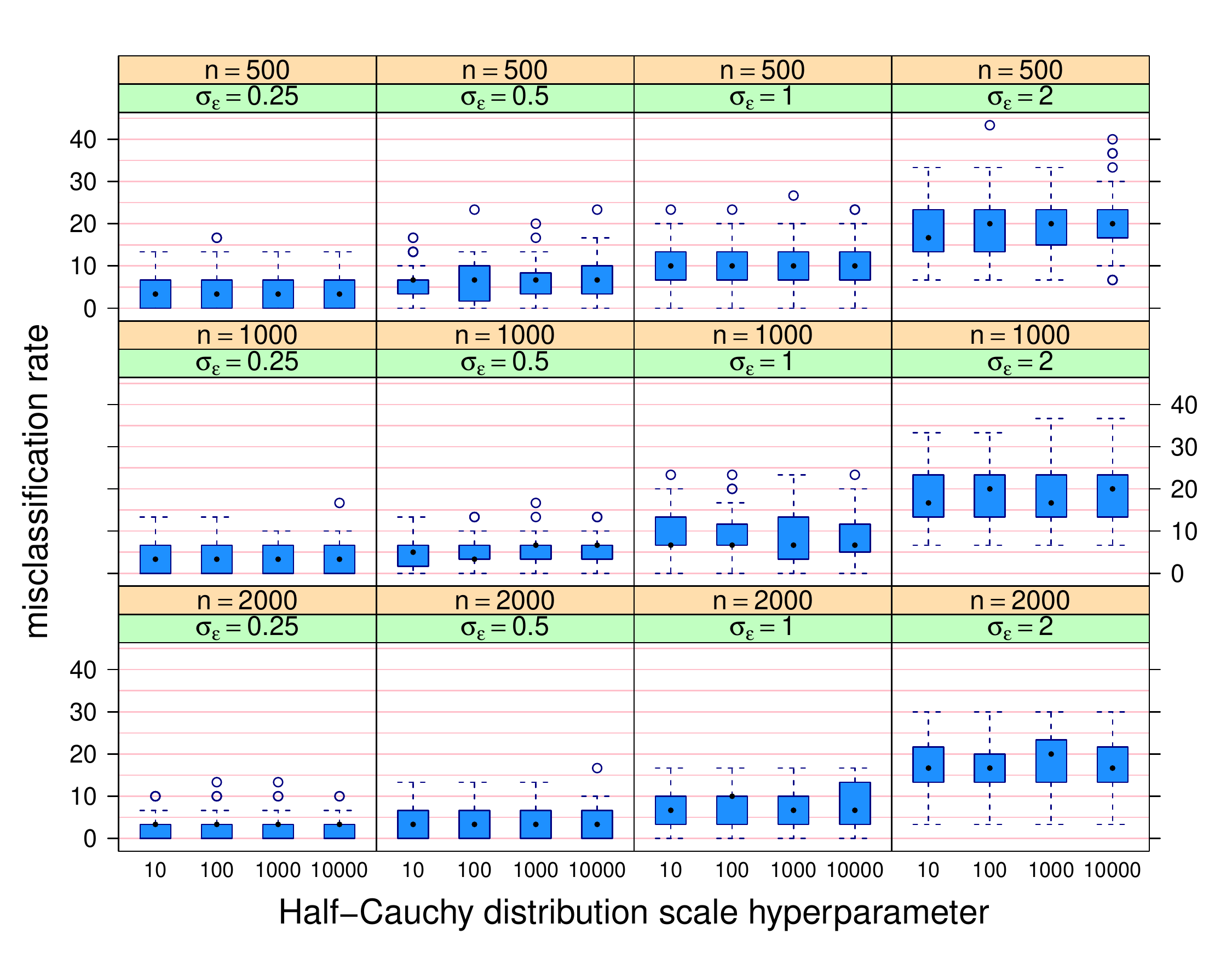}
\caption{
\textit{Side-by-side boxplots of misclassification rate for varying
values of the Half Cauchy distribution scale hyperparameter for the Gaussian response
version of Algorithm \ref{alg:MFVB}, for the first simulation study
described in Section \ref{sec:HypSensChks}.}}
\label{fig:choiceScaHalfCauMFVBgaussCol}
\end{figure}

Figure \ref{fig:choiceScaHalfCauMFVBgaussCol} is similar to 
Figure \ref{fig:choiceScaHalfCauMCMCgaussCol}, but is for
the mean field variational Bayes approach used by 
Algorithm \ref{alg:MFVB} with $\tau$ set to the default 
value of $0.1$. Once again, low sensitivity to the
Half Cauchy distribution scale hyperparameter values
is exhibited.

\vfill\eject

\section*{Reference}

\bib
Hastie, T., Tibshirani, R. \myand Tibshirani, R. (2020). 
Best subset, forward stepwise of lasso? Analysis and
recommendations based on extensive comparisons.
\textit{Statistical Science},\textbf{35}, 579--592.

\end{document}